\def\junk#1{}
\documentclass[10pt, conference, letterpaper]{IEEEtran}

\ifCLASSOPTIONcompsoc
\else
\fi

\ifCLASSINFOpdf
 \usepackage{times}
  \usepackage{caption,color}
  \usepackage{subcaption}
  \usepackage[pdftex]{graphicx}
\else
\fi

\usepackage[cmex10]{amsmath}
\usepackage{amsthm}

\usepackage[linesnumbered,ruled,vlined]{algorithm2e}
\usepackage{multirow}
\usepackage{soul}
\newtheorem{example}{Example}

\newcommand{\superscript}[1]{\ensuremath{^{\textrm{#1}}}}

\begin{document}
\title{A General Constrained Shortest Path Approach for Virtual Path Embedding
\vspace{-4mm}
}

\author{Dmitrii Chemodanov\superscript{*}, Prasad Calyam\superscript{*}, Flavio Esposito\superscript{*}, Andrei Sukhov\superscript{$\dagger$}\\
\small{\superscript{*}University of Missouri-Columbia, USA; \superscript{$\dagger$}Samara State Aerospace University, Russia;} \\ \small{Email: \textit{\{dycbt4, calyamp, espositof\}@missouri.edu, amskh@yandex.ru}}\\
\\
\textit{\large{Technical Report}}

\vspace{-4mm}
}

\maketitle

\begin{abstract}
Network virtualization has become a fundamental technology to deliver services for emerging data-intensive applications in fields such as bioinformatics and retail analytics hosted at multi-data center scales. To create and maintain a successful virtual network service, the problem of generating a constrained path manifests both in the management plane with a physical path creation (chains of virtual network functions or virtual link embedding) and in the data plane with on-demand path adaptation (traffic steering with Service Level Objective (SLO) guarantees). In this paper, 
we define the \textit{virtual path embedding problem} to subsume the virtual link embedding and the constrained traffic steering problems, and propose a new scheme to solve it optimally. Specifically, we introduce a novel algorithm viz., `Neighborhood Method' (NM) which provides an on-demand path with SLO guarantees while reducing expensive over provisioning. 
We show that by solving the Virtual Path Embedding problem in a set of diverse topology scenarios we gain up to $20\%$ in network utilization, and up to $150\%$ in energy efficiency, compared to the existing path embedding solutions.
\end{abstract}


\maketitle

\IEEEdisplaynotcompsoctitleabstractindextext

\IEEEpeerreviewmaketitle

\section{Introduction}
\label{Introduction}

The advent of network virtualization has enabled new business models allowing service providers to share or lease their physical network infrastructure. Even the research community has leveraged such technology offerings within wide-area virtual network federated testbeds such as the Global Environment for Network Innovations (GENI)~\cite{geni} and user communities such as High Energy Physics~\cite{drth}. The virtual network service offerings should meet application Service Level Objectives (SLO) demands on shared (constrained) physical networks. 
The Service Level Objectives are the technical constraints of a Service Level Agreement (SLA) contract. Although the term SLO may include any technical agreement constraint, in this paper we restrict our focus to the constraints within a virtual network service as well as the policies that drive the path of a Virtual Network Functions (NFV) Service chain~\cite{NFVspec}. 
Examples of such SLO demands include guaranteed bandwidth, high reliability, or low latency, while examples of virtual network services include high-performance computing applications such as data-intensive clusters for bioinformatics, or retail analytics hosted at multiple data centers. 

In order to host such virtual network services, infrastructure and cloud providers are required to run a fundamental management protocol commonly referred to as Virtual Network Embedding (VNE)~\cite{Esposito:2013:SES, VNE-survey}. The VNE is the NP-hard graph matching problem of mapping a constrained virtual network on top of a shared physical network. 
Once a virtual network has been embedded, traffic steering techniques need to be enforced to avoid SLO violations during both the virtual network creation phase (within the management plane), as well as when a network service is already functional (within the data plane). Moreover, to connect the (virtual) instances of a chain of virtual network functions, many middleboxes need to be connected with different traffic policies (path constraints). 

In essence, to create and maintain a successful virtual network service, the problem of generating a constrained path is fundamental both in the management plane with a physical path creation (NFV chain instantiation or virtual link embedding) and in the data plane with on-demand path adaptation (traffic steering with SLO guarantees).
In this paper, we generalize the NFV chain instantiation, the virtual link embedding and the SLO-constrained traffic steering problems, and subsume them as a unique {\it virtual path embedding} problem. The Virtual Path Embedding problem is the problem of embedding a virtual path on a physical or logical constrained loop-free path minimizing the network over provisioning.
Our notion of over provisioning (that hinders infrastructure providers' revenue maximization) can be understood in a case where three switches are used for maintaining one virtual link (point-to-point path) when there is a solution alternative that needs only two switches. We define a path to be optimal if it satisfies all SLO constraints for the minimum number of hops. Other metrics such as bandwidth, delay and jitter directly reflect a service quality, and have to be declared in a SLO. 

Solutions to particular instances of the virtual path embedding problem already exist in both management and data planes. Let us consider $e.g.$, the multi-constrained path resource reservation problem, $i.e.$, the link embedding phase of the VNE problem. After the source and destination physical nodes have been identified, most of the link embedding solutions (see~\cite{Esposito:2013:SES, VNE-survey} for a survey) seek the reservation of a loop-free physical path applying the $k$-shortest path algorithm~\cite{eppstein}, often with $k=1$.
Even when node and link embedding are considered jointly, heuristics do not guarantee that an optimal loop-free physical path will be found, even if a feasible embedding solution exists~\cite{nodeembedding}. This in turn leads to lower physical network utilization, suboptimal path allocation and consequent energy wastage caused by over provisioning.

Existing data plane traffic steering solutions also have limitations~\cite{QoSRoutingSurvey}.
Some heuristics solve this NP-complete problem~\cite{ktmk} considering only {\it additive/multiplicative or path constraints}~\cite{yuli}, $e.g.$, delay and jitter, while others consider {\it concave, min/max or link} constraints, such as bandwidth~\cite{net_utilization}. Even when such solutions consider both constraint types, they are often suboptimal, $e.g.$, they may find longer paths~\cite{yuli,jaffe}, or they are not guaranteed to find a feasible solution even if it exists~\cite{jsmr}.

\noindent
{\bf Our Contributions.} In this paper, we present a general multi-constrained optimal (virtual) path finder method for the virtual path embedding problem, $i.e.,$ for both physical path creation (NFV chains and virtual link embedding) applications, and for on-demand path adaptation (traffic steering) techniques $i.e.$, after a virtual network has been created.
Our  algorithm $viz.$, {\it ``Neighborhoods Method''} (NM), provides on-demand virtual path embedding guarantees while reducing over provisioning, and runs in polynomial time when accepting multiple link and a single path constraints.

When instead we seek paths with multiple link and multiple path constraints, our NM algorithm runs in exponential time and hence it is intractable for large-scale physical networks~\cite{widyono}.
To show our approach benefits, we analyze the complexity of NM, and compare it with two related solutions: the Exhaustive Breadth-First Search (EBFS)~\cite{widyono} and the extended version of Dijkstra (EDijkstra)~\cite{wacr} algorithms.  
We also show how a path with similar constraints can be found only in exponential time if we use the EBFS algorithm instead. 

We evaluate our NM by first showing how beneficial the approach is in terms of physical network utilization when applied to existing management plane (VNE) solutions. Then we show NM is beneficial for data plane SLO-adhering solutions such as on-demand traffic steering.
For every tested VNE solution, we found that the number of embedded VN requests (and thereby the providers' revenue) increase when using NM as link embedding.
In particular, with NM we were able to allocate at best twice as many virtual links.
To evaluate our proposal with respect to path adaptation data plane solutions, we compare  instead NM against Extended-Dijkstra (EDijkstra) a well-known technique that combines links pruning phase with Dijkstra algorithm.~\cite{wacr}. Our  simulation results show that our approach leads to gains of up to $20\%$ in physical network utilization, and up to $150\%$ in energy efficiency.

The rest of the paper is organized as follows: In Section~\ref{related_works}, we discuss a few virtual path embedding solutions pertaining to VNE and the multi-constrained path problem. 
In Section~\ref{nm}, we present details of our NM approach. 
Section~\ref{implementation} describes our evaluation methodology, performance metrics and results which show NM's effectiveness. Section~\ref{conclusion} concludes the paper.

\section{Related Work}
\label{related_works}

Our NM approach has general applicability to any instance of the virtual path embedding problem.
In this section we relate our contributions to a limited set of path embedding strategies which focus either on the VNE management protocol, or on current traffic steering approaches for dynamic path adaptation in a data plane. 
The literature of both sub-areas is vast, and we only focus here on a few relevant papers which help us highlight our contributions. A complete survey of recent VNE solutions and NFV is discussed in~\cite{Esposito:2013:SES, VNE-survey,nfv_survey, nfv_survey1}, while a survey on multi-constrained path solutions can be found in~\cite{QoSRoutingSurvey}.

\noindent
{\bf Virtual link embedding solutions.} 
The VNE problem requires a constrained virtual network to be mapped on top of a physical network hosted by a single infrastructure provider, or by a federation of them. To solve this NP-hard~\cite{chva}  problem, researchers have proposed centralized~\cite{yyrc, zham,joaz,lika} and distributed~\cite{nodeembedding,polyvine,HS} heuristics that either separate the node embedding from the link embedding phase~\cite{nodeembedding,yyrc,polyvine,HS}, or simultaneously apply the two phases~\cite{lika}.
Most of the (centralized or distributed) VNE solutions which separate the node from the link embedding use a $k$-shortest path algorithm, often with $k =1$.
A link embedding based on a $k$-shortest path may be suboptimal. 
The  widely cited approach used by Yu $et$ $al.$~\cite{yyrc} for example, begins a $k$-shortest path embedding with $k=1$, and then increases $k$ until a path which satisfies all constraints is found. Such schema leads to an exhaustive search with recalculation of previously found paths at each iteration, while our NM finds the optimal paths if they exist with just a single pass. 
In Lischka $et$ $al.$~\cite{lika}, the authors propose a VNE based on subgraph isomorphism detection of the original physical network graph. The link suboptimality in this case arises from the physical paths having  a length lower than a predefined heuristically chosen value. The heuristic is necessary to reduce the search space that is otherwise exponential.
Our NM approach applied to the link embedding is agnostic to the type of heuristic used to solve the NP-hard node embedding problem. %
To show that this is the case, in our evaluation section we compare a representative set of existing distributed VNE solutions; replacing the $k$-shortest path algorithm with our NM leads to a higher virtual network request acceptance rate, and therefore a higher infrastructure provider revenue, with slightly higher time to solution.

\noindent
{\bf Multi-constrained Path Solutions.}
The problem of providing a path with multiple (SLO) constraints is NP-complete~\cite{jaffe}, and its complexity has inspired many heuristics. 
Most of these heuristics group multiple metrics into a single function to reduce the problem to a single constrained routing problem~\cite{ktmk}, and then solve the routing optimization problem using $e.g.$, Lagrangian relaxation~\cite{jsmr}. 
Depending on the constraints, these approaches may not reach an optimal route (while ours does), $i.e.$, the duality gap introduced by the Lagrangian function may not be zero.  Other suboptimal solutions use a $k$-constrained approach. In~\cite{yuli} for example, Yuli $et$ $al.$ use $k$ path constraints and attempt to find a solution in $O(|V|^2|E|)$, where $V$ and $E$ are the number of vertices and edges of the physical network graph, respectively. NM's running time is comparable with the above solutions in multiple link and a single path constraints case, but NM is exact and optimal.

The non-heuristic solution proposed by Jaffe $et$ $al$~\cite{jaffe} offer a distributed path finder solution for a two-path constraint problem with $O(|V|^5blog(|V|b))$, where $b$ is the largest weight of all links in the substrate network which makes the algorithm pseudo-polynomial.
Wang $et$ $al.$~\cite{wacr} leverage the Dijkstra shortest path algorithm to propose a bandwidth-delay constraint routing approach  called ``pre-routing schema". During such pre-routing, their algorithm excludes links 
with unfeasible bandwidth constraints. 
Their algorithm provides a solution in $O(|V|log|V| + |E|)$.  Despite EDijkstra runs in polynomial time it has to omit path length optimization due to delay constraint satisfaction, whereas our NM algorithm allows such optimization. 
In~\cite{widyono}, the authors propose a ``Constrained Bellman-Ford'' which leverages EBFS. The algorithm finds (in exponential time) a path that satisfies constraints on delay, while simultaneously minimizing a broader notion of cost. 
\textcolor{black}{The authors in~\cite{exact_qos} propose an exact algorithm for the NP-hard multiple path constraints problem, and apply several search space reduction techniques for the $k$-shortest path algorithm resulting in a suboptimal pseudo-polynomial or even exponential solution.}
Hop-count optimization techniques have been floated before, $e.g.$, to QoS flow scheduling where bandwidth constraint are also taking into account~\cite{net_utilization}. 
\textcolor{black}{NM instead finds the optimal hop-count (virtual) paths to solve the virtual path embedding problem.}
\section{Neighborhood Method}

\label{nm}
\theoremstyle{plain}
\newtheorem{th_nm}{Theorem}
\newtheorem{col_nm}{Corollary}

\noindent
{\bf Problem definition.} The Virtual Path Embedding problem is the problem of embedding a virtual path on a physical or logical constrained loop-free path minimizing the network over provisioning.
To solve this general problem with $l$ link constraints and $p$ path constraints, we devised the Neighborhoods Method.
We define a path to be optimal if it is the shortest (in terms of number of hops) that satisfies our SLO constraints hence minimizing the network over-provisioning.

\noindent
{\bf Why the name ``Neighborhoods Method"?} Most path seeking algorithms require at least two inputs: $(i)$ knowledge of neighbors, and $(ii)$ awareness of all adjacent link costs, often dictated by SLO constraints or policies. Such constraints are then used by the path seeking algorithm to compute the lowest-cost paths. The Dijkstra algorithm for example finds the shortest path traversing the source neighbors and the neighbors of their neighbors, recursively. This recursive notion leads to our definition of ``neighborhoods'' in NM, $i.e$,  a set of nodes that can be reached from the source node with the same number of hops, where each ``neighborhood'' contains unique elements.

\begin{figure}
\centering
\includegraphics[width=0.95\linewidth]{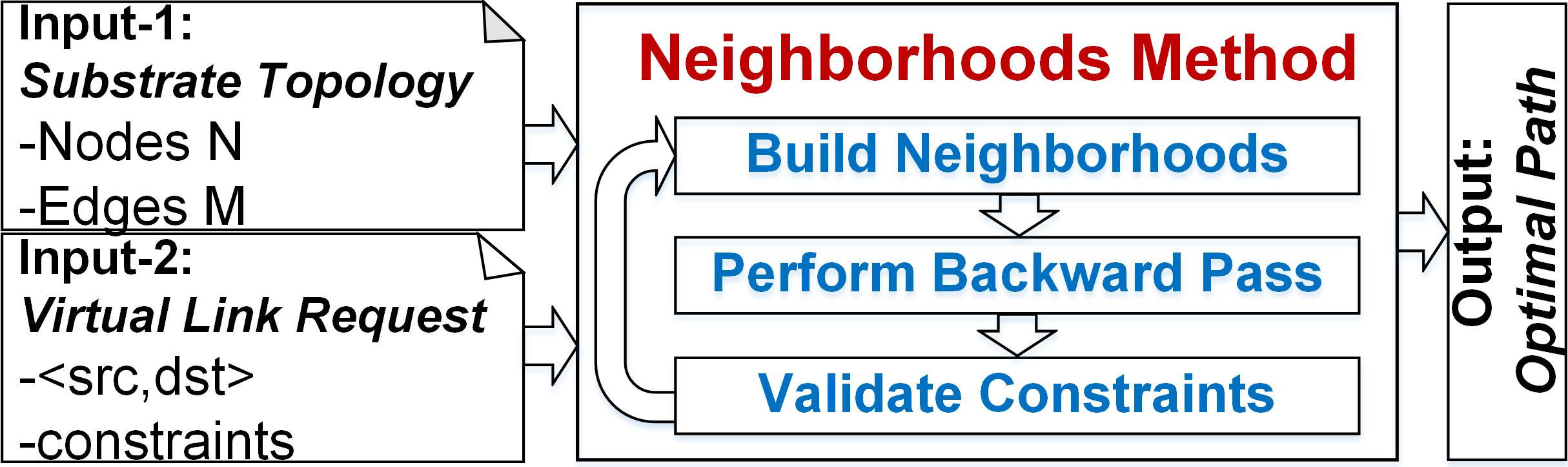}
\caption{\footnotesize{Neighborhood Method Workflow in $l \oplus p$ case: The $k^{th}$-hop Neighborhood  Build (forward pass) first, and the Backward Phase later find the best loop-free constrained path if it exists. If a third phase yields unfeasibility, the forward pass is repeated recursively with a larger ($k+1$) neighborhood.}}
\label{general_diagram}
\end{figure}

\subsection{The NM General Case ($l\oplus p$ case)} 
The general case of our NM algorithm accepts multiple links and multiple path constraints, it has exponential computational complexity~\cite{jaffe}, and scales poorly to multi data center virtual network service applications. 
Note however, that existing heuristics such as single mixed metric~\cite{ktmk} or Lagrangian relaxation~\cite{jsmr} can reduce the $l\oplus p$ to a $l\oplus1$ case.

Figure$~\ref{general_diagram}$ and Algorithm~\ref{genAlg} show the workflow of the general NM algorithm. NM is executed in three steps: $(i)$ the \textit{forward pass} or neighborhoods building,  $(ii)$ a \textit{backward pass}, and a final $(iii)$ \textit{back track path validation} step.
During the forward pass, NM builds the neighborhoods to estimate the path length. The backward pass is used to find end-to-end paths with a given length and the final constraints validation step is used to decide whether or not the path search should be extended to longer path candidates involving more neighbors. 

\begin{algorithm}[h]

\footnotesize
\KwIn{$X$:= src $Y$:= dest, $C$:= constraints list (links and paths)}
\KwOut{Shortest $Route$ between $X$ and $Y$ satisfing $C$}

\SetAlgoLined
\Begin{
\tcc{Build neighborhoods $<NH>$ from X to Y}
$<NH> \longleftarrow Build$ $Neighborhoods$\

\tcc{$Backward$ $Pass$ to find whole set of routes <Route> with length equal  $<NH>$ size}

$<Route> \longleftarrow Perform$ $Backward$ $Pass$\

\tcc{Validate QoS constraints; if not satisfied, add one neighborhood and repeat $Backward$ $Pass$}

$Validate$ $Constraints$\
}
\caption{General NM algorithm ($l\oplus p$ case)}
\label{genAlg}
\end{algorithm}

\begin{figure}[h!]
\centering
\begin{subfigure}[t]{0.12\textwidth}
\centering
\includegraphics[width=1\linewidth]{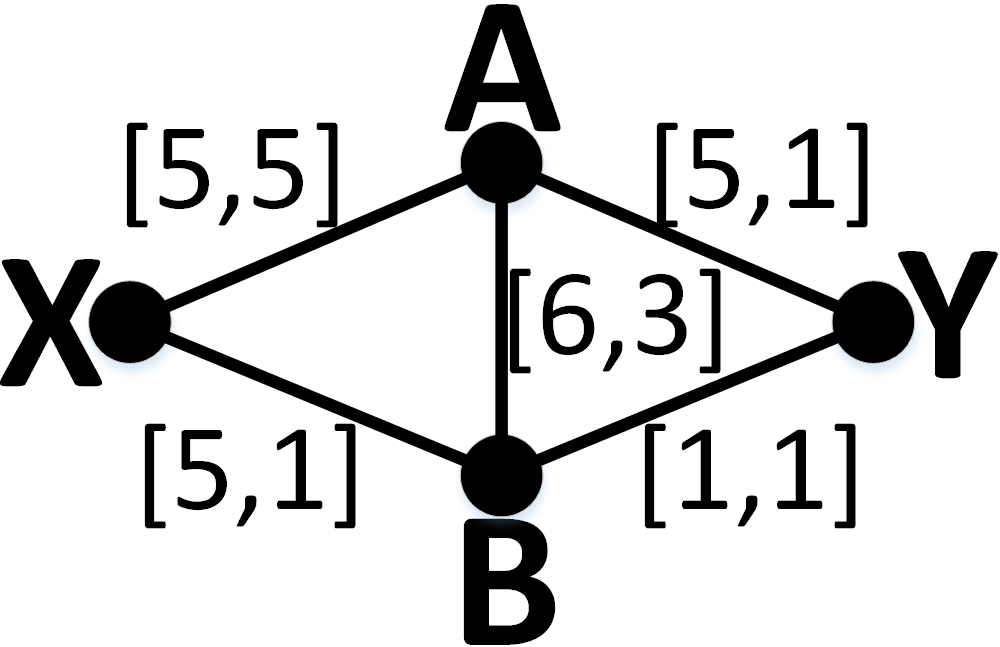}
\caption{}
\label{example_n}
\end{subfigure}
~
\begin{subfigure}[t]{0.12\textwidth}
\centering
\includegraphics[width=1\linewidth]{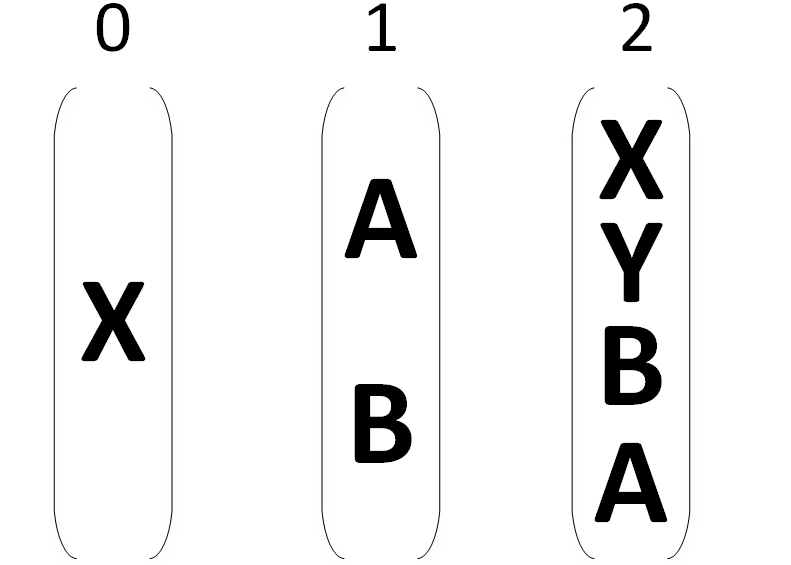}
\caption{}
\label{example_f1}
\end{subfigure}
~
\begin{subfigure}[t]{0.12\textwidth}
\centering
\includegraphics[width=1\linewidth]{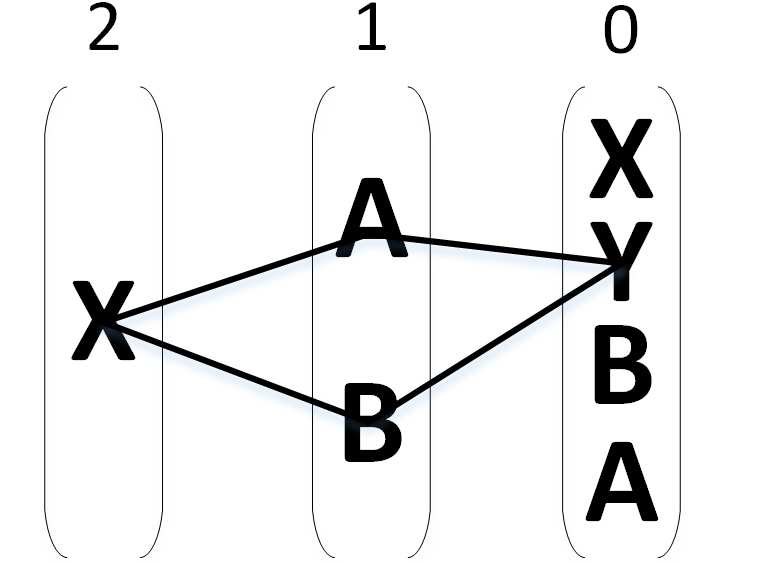}
\caption{}
\label{example_b1}
\end{subfigure}

\begin{subfigure}[t]{0.14\textwidth}
\centering
\includegraphics[width=1\linewidth]{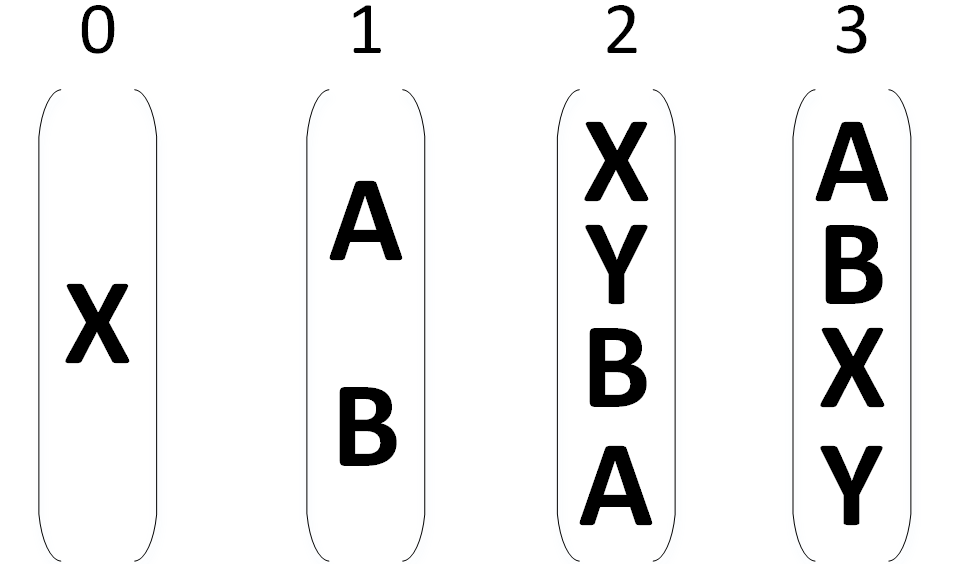}
\caption{}
\label{example_f2}
\end{subfigure}
~
\begin{subfigure}[t]{0.14\textwidth}
\centering
\includegraphics[width=1\linewidth]{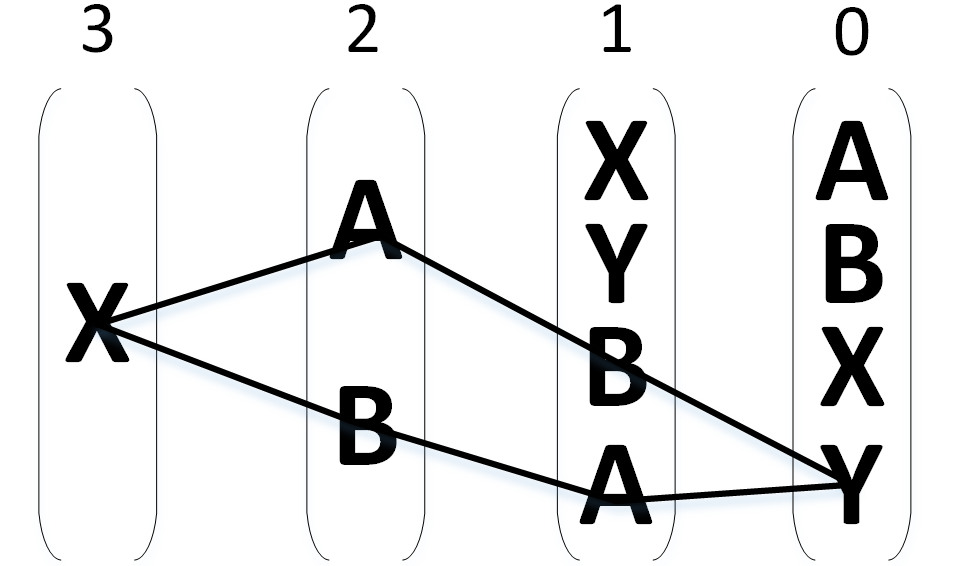}
\caption{}
\label{example_b2}
\end{subfigure}
\caption{\footnotesize{Illustrative example of a NM run in the general case (when $l$ link and $p$ path constraints are specified: (a) simple network configuration with [bandwidth, delay] constraints for each link; (b) forward pass finds the shortest length to Y during first iteration; (c) backward pass identifies all the shortest paths candidates; (d) forward pass finds the shortest + 1 length to Y during second iteration;(e) backward pass identifies all the shortest + 1 paths candidates.}}
\label{example_case3}
\end{figure}

\begin{example}\label{example0}
To illustrate how our NM algorithm works in the $l\oplus p$ case, consider an example network consisting of 4 nodes $X$, $Y$, $A$ and $B$ (Figure$~\ref{example_n}$). Each link has two metrics - a link metric (first value) $e.g.$, bandwidth $bw$ and a path metric (second value) $e.g.$, propagation $delay$. 
Our aim is to find a route from $X$ to $Y$ which satisfies constraints, e.g.  $bw \geq 5$ and $delay \leq 5$. In the first step, we build neighborhoods starting from source node $X$ until destination node $Y$ is reached as shown in Figure~\ref{example_f1}. As soon as we reach $Y$, we begin the second step where we perform backward pass to find the full set of shortest routes as shown in Figure~\ref{example_b1}. If a route with satisfying cost among the shortest ones is found, NM will stop searching and returns this route. In our case, there is no route which satisfies all constraints among the shortest ones. If an appropriate route is not found, the NM builds an additional neighborhood as shown in Figure~\ref{example_f2}. Following this, NM performs backward pass and finds all the shortest + 1 length routes as shown in Figure~\ref{example_b2}. In the last step, NM looks for appropriate solution among found the shortest + 1 length routes, and if it is not there NM terminates due to the maximum path length violation. In our case, $X\rightarrow B\rightarrow A\rightarrow Y$ satisfies all constraints, and NM returns this solution.
\end{example}

\noindent
{\bf Forward Pass for $l\oplus p$.}
In the first step, NM successively builds $<NH>$ from the source node $X$ to reach the destination node $Y$. Algorithm~\ref{buildNb} describes in detail the forward pass of NM. 
To build a $NH$, we add therein neighbors (adjacent nodes) of each node $n$ in $cNH$ (line 6). For example, the first $NH$ includes nearest neighbors of the source node $X$ (which is in the zero $NH$), and the second $NH$ contains nearest neighbors of nodes in the first $NH$, and so on. The first step ends as soon as the destination node $Y$ appears in the $cNH$ (line 4), or a number of the $<NH>$ is more or equal to the number of nodes (line 9).
 
\begin{algorithm}[h]
\footnotesize
\tcc{Returns: neighborhoods list $<NH>$ from $X$ to $Y$ if reachable}
\KwIn{$X$:= src, $Y$:= dest}
\KwOut{$<NH>$ from $X$ to $Y$}

\SetAlgoLined
\Begin{
$cNH \longleftarrow X$\\
$<NH> \longleftarrow <NH> \cup cNH$\\
\While{$Y \notin cNH$} 
{
$NH \longleftarrow \emptyset$\\
\ForEach{Node $n \in cNH$}{$NH \longleftarrow NH$ $\cup$ neighbors of $n$ } 
\eIf{$<NH>.size \leq$ number of nodes}
{
$<NH> \longleftarrow <NH> \cup NH$\\
$cNH \longleftarrow NH$
}{$Y$ is unreachable}
}
} 
\caption{Build Neighborhoods ($l\oplus p$ case)}
\label{buildNb}
\end{algorithm}

\noindent
{\bf Backward Pass for $l \oplus p$.}
Algorithm~\ref{backwardAlg} details the second step of NM - backward pass. 
This step works similar to the Breadth-First Search method with the only difference being that we do not process all neighbors of each node $n$ but only those which are within previous $NH$ (line 10). The first step is to find the intersection between neighbors of the destination node $Y$ with its previous $NH$ (line 5). This intersection is not an empty set, it contains at least one node $nh$. For all obtained nodes we again build the intersection of their neighbors with the penultimate $NH$ (line 16). The second step ends as soon as we hit the zero $NH$ (line 6), and as a result we obtain the collection of all paths with a length of $<NH>$ size between the source node $X$ and the destination node $Y$. \textit{A distinctive feature of our NM is the construction of the intersection of two neighborhoods, one of which will always be the nearest one. Thus, our NM approach greatly reduces the number of comparisons i.e., the computational complexity of the analysis.}

\begin{algorithm}[h]
\footnotesize
\KwIn{$<NH>$ - list of the sets of nodes, $l$ - link constraints}
\KwOut{All possible paths $<path>$ between $X$ and $Y$ of a given length equal to hte size of $<NH>$} 

\SetAlgoLined
\Begin{
$path \longleftarrow Y$\\
$<path> \longleftarrow <path> \cup path$\\
$k \longleftarrow 1$\\
$NH \longleftarrow <NH>[size-k]$\\
\While{$NH$ $\neq <NH>[0]$}
{
$<tempPath> \longleftarrow \emptyset$\\
\ForEach{$path \in <path>$}
{
Node $n \longleftarrow path[1]$\\
\ForEach{Neighbor $nh \in$ neighbors of $n \cap NH$}{$<tempPath>\longleftarrow neighbor \cup path$}
}
$<path> \longleftarrow <tempPath>$ \\
$k \longleftarrow k+1$\\
$NH \longleftarrow NH[size-k]$\\
}
} 
\caption{Perform Backward Pass ($l\oplus p$ case)}
\label{backwardAlg}
\end{algorithm}

\noindent
{\bf Constraints Validation for $l\oplus p$.}
Finally, we check relevant candidates i.e., paths that satisfy required cost among found ones. If appropriate paths are not discovered, we build one more ($N+1$) neighborhood (Figure~\ref{forward}) and repeat backward pass (line 3 of Algorithm~\ref{genAlg}) and subsequently obtain all shortest + 1 length paths. Again, we check that the retrieved routes satisfy all constraints, and if none of them satisfy, we build a neighborhood $N + 2$. Iteration continues until the solution length will be equal to the number of nodes.

\subsection{NM for $l$ links and a single path constraints ($l\oplus1$ case)}
\label{nm_l_1}

If we only have $l$ links and a single path constraint NM runs in polynomial time, as unnecessary iterations are avoided.

\begin{figure}[t!]
\centering
\begin{subfigure}[b]{0.13\textwidth}
\centering
\includegraphics[width=1\linewidth]{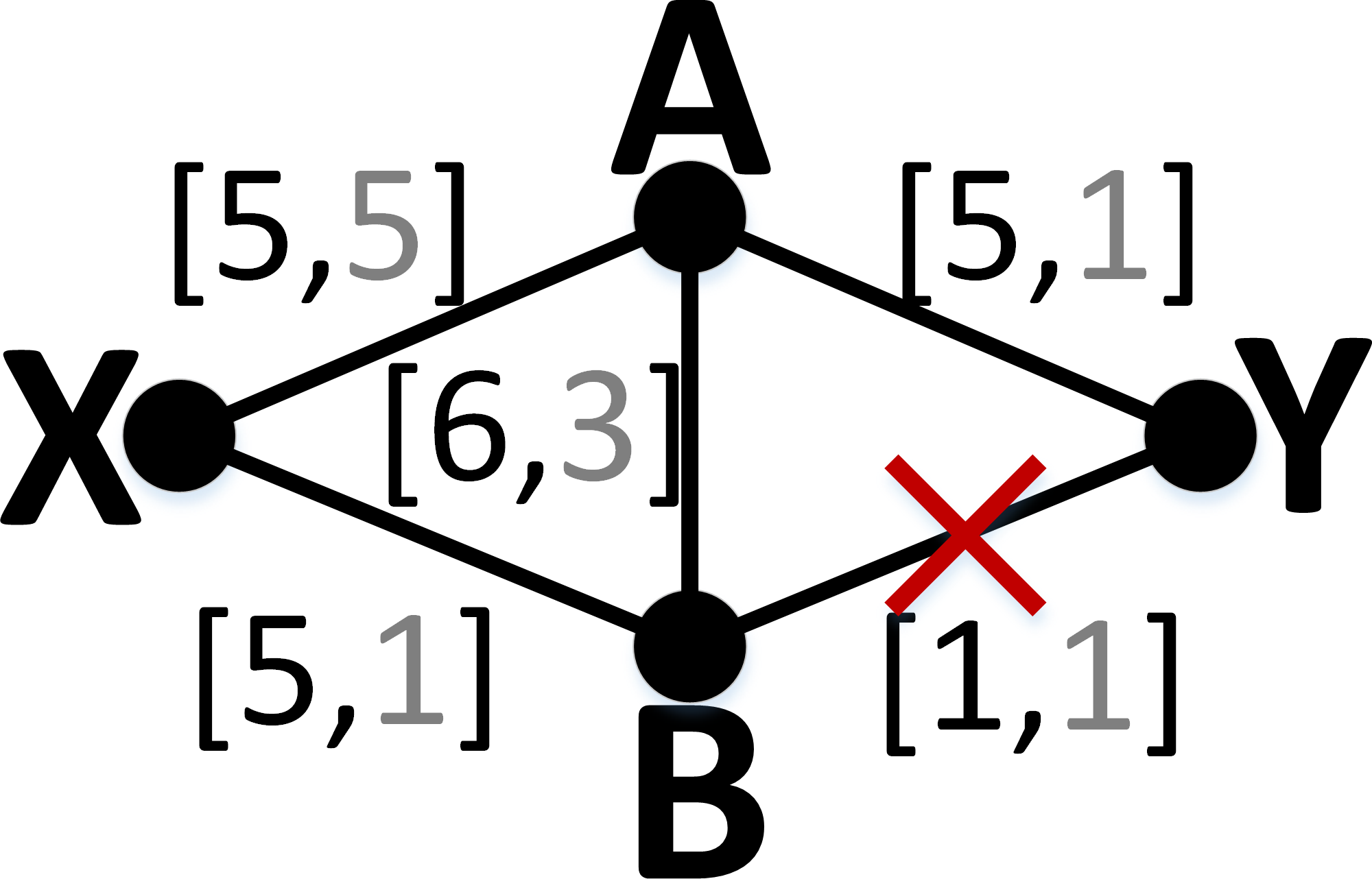}
\caption{}
\label{example_n1}
\end{subfigure}
~
\begin{subfigure}[b]{0.16\textwidth}
\centering
\includegraphics[width=1\linewidth]{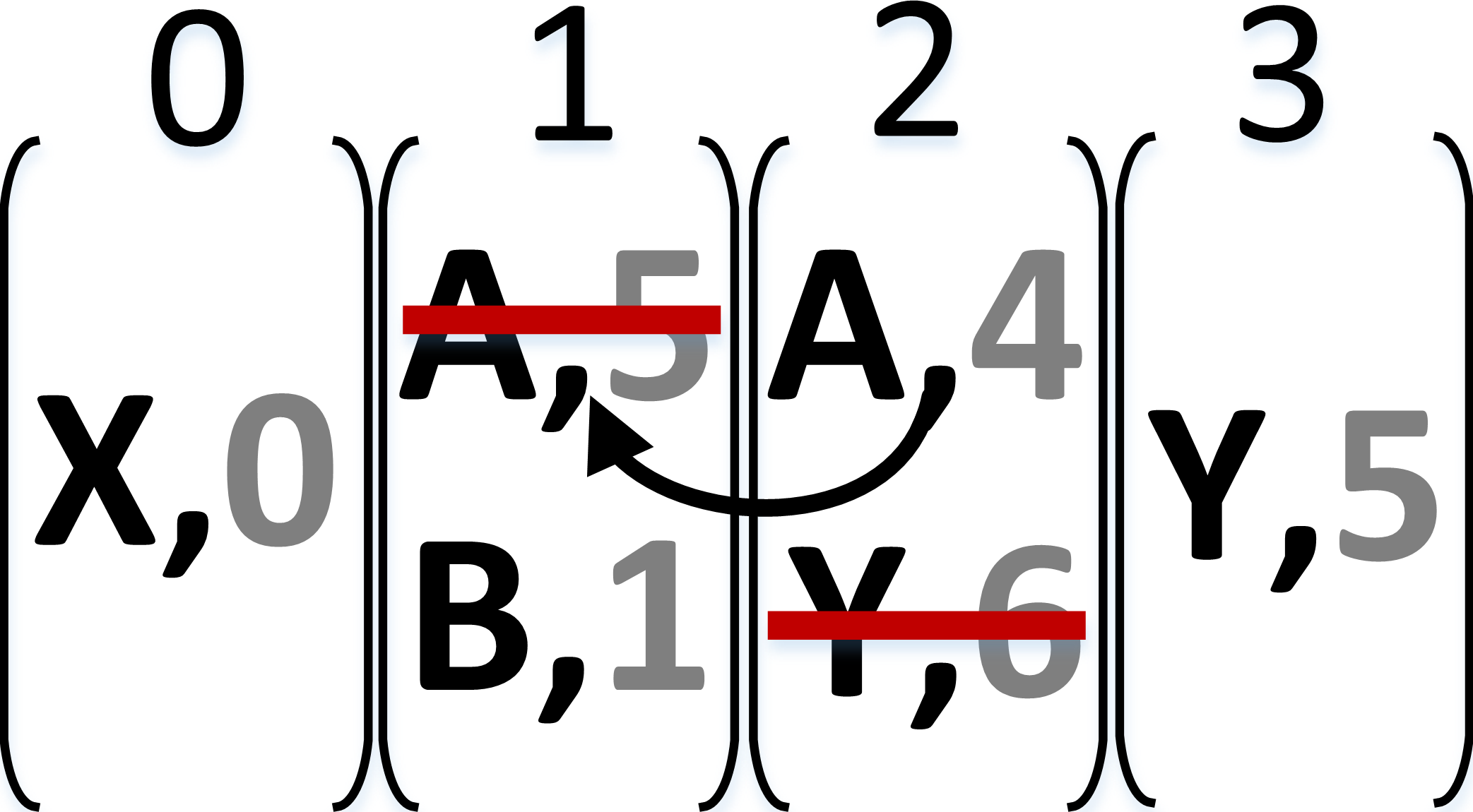}
\caption{}
\label{example_p2}
\end{subfigure}
~
\begin{subfigure}[b]{0.16\textwidth}
\centering
\includegraphics[width=1\linewidth]{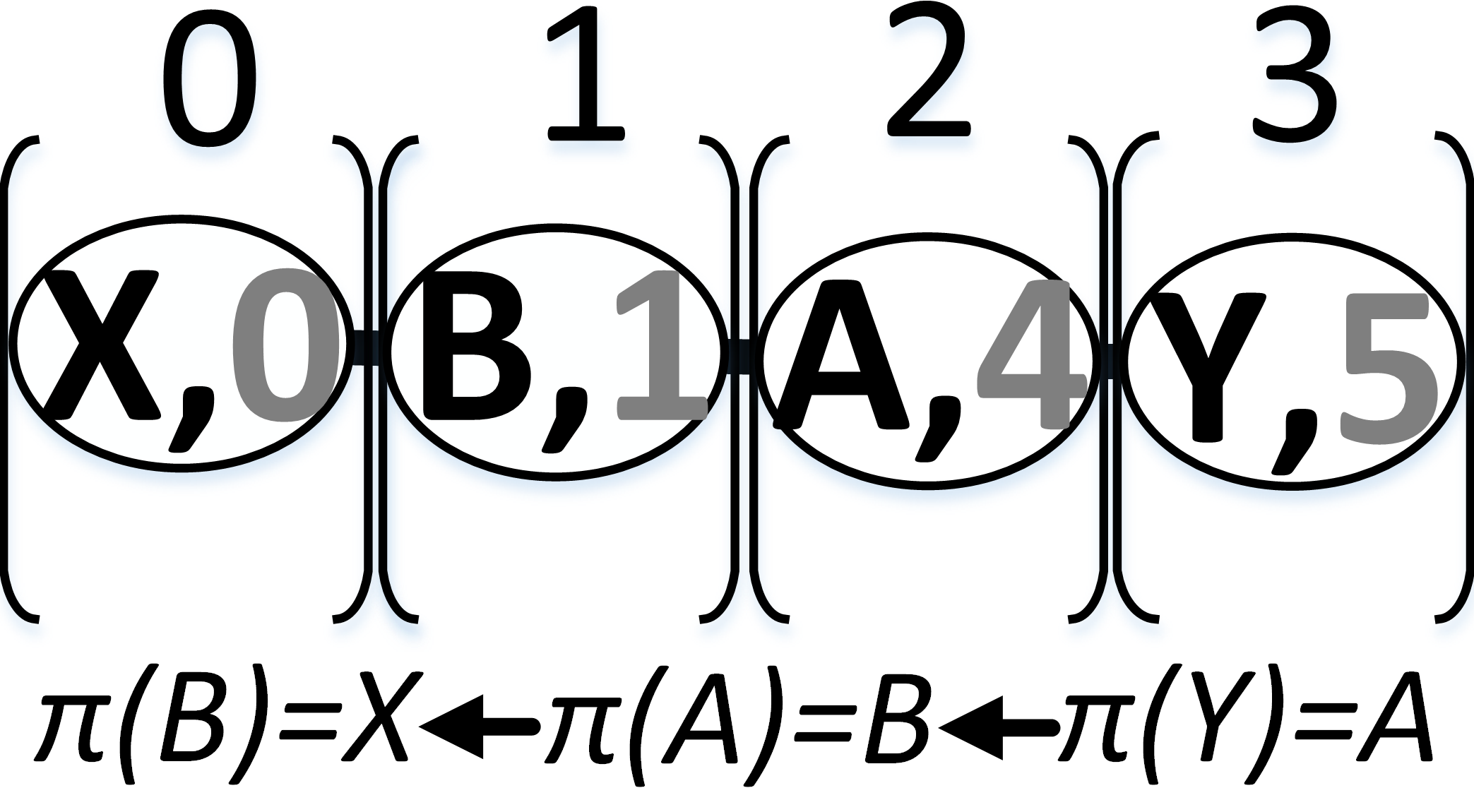}
\caption{}
\label{example_p3}
\end{subfigure}
\caption{\footnotesize{Running example of NM with $l$ links and 1 path constraint specified ($l\oplus1$ case): (a) In the \textit{pre-routing phase} NM prunes $B\rightarrow Y$ due a $bw$ violation; (b) In the \textit{forward phase} NM finds the best length to Y discarding previous results if better subpaths are found; (c) \textit{back track phase} identifies the valid path X, B, A, Y by recursive predecessor visits.}}
\label{example_case}
\end{figure}

\begin{example}
Consider Figure$~\ref{example_n1}$. 
On the link $(X,A)$, the first value $5$ refers to a link constraint in this case bandwidth, and the second $5$ refers to a path metric, in this case end-to-end delay. NM finds a path from $X$ to $Y$ satisfying the two constraints $bw \geq 5$ and $delay \leq 5$ as follows: In the \textit{pre-routing phase} (Figure$~\ref{example_n}$) In the \textit{pre-routing phase} NM prunes $B\rightarrow Y$ due a $bw$ violation. In the \textit{forward phase} (Figure$~\ref{example_p2}$) NM finds the best length to Y discarding previous results if better subpaths are found.
During the  \textit{back trace phase} we then recursively construct the actual optimal path using the node identifiers found in the forward phase (Figure$~\ref{example_p3}$). 
\end{example}

\noindent
{\bf NM pseudocode for $l \oplus 1$}. Algorithm~\ref{buildNbWithPath} describes the NM algorithm for the $l\oplus1$ case with additive path metric. Note that multiplicative constraints can be converted to additive by composing them with a logarithmic function.
Each node $u$ contains information about its predecessor $\pi(u)$, the distance $D(u)$ and the location of its last neighborhood $L_{NH}(u)$. Note that we do not delete empty neighborhoods to be able detect a negative weight cycle (line 33).
During the pre-routing phase, $\pi(u)$ and $L_{NH}(u)$ are set to $NIL$, whereas $D(u)$ is set to $0$ (line 3-5). We then set all edge weights that do not satisfy $l$ constraints to $\infty$  to avoid using these links during the forward phase (line 19).

During forward phase we successively build the neighborhoods $<NH>$ , where for each {\it current} neighborhood $cNH$ of $u$ we exclude all neighbors $v$ that do not satisfy the path constraint  or that have previous distance lower than $dist$ (line 31). In addition, we remove neighbor $v$ from the previous neighborhood location $L_{NH}(v)$ (line 24) and \textcolor{black}{save a new neighborhood location of $v$} (line 25).
The forward phase ends as the destination node $Y$ appears in the current neighborhood $cNH$ (line 4). If  the number of neighborhoods $<NH>$ is equal to the nodes number (line 33) we terminate the algorithm concluding that a negative weight cycle is detected, or that a path does not exist (line 36).
During the back track phase we recursively find the shortest possible path between $X$ and $Y$ which satisfies $l\oplus1$ (line 41).

\begin{algorithm}[h]
\footnotesize
\KwIn{$X$:= src, $Y$:= dest, $l$ link constraints, $p$  single path constraint.}
\KwOut{The shortest possible $path$ between $X$ and $Y$ which satisfies $l\oplus1$.} 
\SetAlgoLined
\Begin{
\tcc{\textbf{pre-routing phase:}}
\ForEach{Vertex $u \in |V|$}
{
$\pi(u) \leftarrow NIL$\\
$D(u) \leftarrow 0$\\
$L_{NH}(u) \leftarrow NIL$\\
}
\ForEach{Edge $u\rightarrow v \in |E|$}
{
\If{$u\rightarrow v$ does not satisfy $l$ }
{
$w(u\rightarrow v)\leftarrow \infty$\\
}
}

\tcc{\textbf{forwarding phase:}}
$cNH \longleftarrow X$\\
$<NH> \longleftarrow <NH> \cup cNH$\\
\While{$Y \notin cNH$} 
{
$NH \longleftarrow \emptyset$\\
\ForEach{Vertex $u \in cNH$}
{
\ForEach{Neighbor $v \in adjacent(u)$ }
{
$dist \leftarrow D(u) + w(u\rightarrow v)$\\
\If{$dist < D(v)$ and $dist<p$} 
{
$\pi(v) \leftarrow u$\\
$D(v) \leftarrow dist$\\
$NH \longleftarrow NH \cup (v)$\\
\If{$L_{NH}(v)!=NIL$}
{
${<NH>[L_{NH}(v)]} \leftarrow {<NH>[L_{NH}(v)]} - v$\\
$L_{NH}(v) \leftarrow <NH>.size + 1$\\
}
}
}
} 
\uIf{$NH \notin \emptyset$ and $<NH>.size < |V|$}
{
$<NH> \leftarrow <NH> \cup NH$\\
$cNH \leftarrow NH$\\
}
\uElseIf{$<NH>.size == |V|$}
{Return Negative Weight Cycle is detected. }
\Else{Return $Y$ is unreachable. }
}

\tcc{\textbf{back track phase:}}
$path \leftarrow Y$\\
\While{$\pi(path(1)) \neq NIL$}
{
$path \leftarrow \pi(path(1)) \cup path$\\
}
}
\caption{NM in $l\oplus1$ case}
\label{buildNbWithPath}
\end{algorithm}

\subsection{NM Optimal Solution and Constraints Satisfaction}
\label{generalTh}

\begin{th_nm}
\emph{(Theorem of the Optimal Solution)}
\label{th1}
\\NM always finds \textit{the optimal path} if it exists.
\end{th_nm}
\junk{
\begin{proof} 
The proof (by contradiction) is included in our technical report~\cite{tech_report}.
\end{proof} 
By contradiction we also proved optimality of our NM method and  present a corollary to show that such solution satisfies $l\oplus1$ constraints~\cite{tech_report}.
}
\begin{proof} 
To prove this theorem we need to prove: firstly, that the forward pass of the general NM checks constraints satisfaction for each possible path length between the source and the destination nodes; secondly, that the backward pass of NM can return any path of a given length.
The first thesis can be proved by contradiction: assume that there is the 
{\it optimal path} with ($i$) the minimum or ($ii$) the maximum length, and for this length the constraints have not been checked. Firstly, the forward pass step starts building neighborhoods from the first neighborhood, meaning that we check for constraints satisfaction at the path length=1. If there is a path with length lower than $1$, we have the special case in which the source node is also the destination assumed to be satisfied, which in turn contradicts with assumption ($i$). 
Further, the forward pass step continues to build neighborhoods until their number is equal to the number of nodes, or the destination node with satisfied constraints is in the last neighborhood. In the first case, the forward pass ends by checking constraints satisfaction at the maximum possible length of a loop-free solution, and there will never be a case in which a complex path (with loops) can be provided due to a contradiction with \textit{the optimal solution} definition. In the second case, the forward pass ends by finding the minimum length at which all constraints are satisfied, hence NM will provide this path which is by definition \textit{the optimal solution}. In both cases we contradict assumption ($ii$). 
The second thesis can be also proved by contradiction: assume that NM has not found a path with the $N$ hops length between the source and the destination nodes. This is possible only if at least one of the path's nodes has not appeared in the neighborhoods $<NH>$. In this case, it suggests that this node is not accessible from the source node within $N$ hops, i.e., if it is unreachable, or path does not have the $N$ hops length, which contradicts our assumption.
In summary, we proved that the forward pass of NM checks constraints satisfaction at each possible path length between the source and the destination nodes. At the same time, the backward pass can return any path of a given length which provided by the forward pass. 
Consequently, we can conclude that - if  an \textit{optimal solution} exists, then NM will find it.
\end{proof}

\begin{col_nm}
\emph{(The $l\oplus1$ Constraints Satisfaction)}
\label{th3}
\\NM always provides a solution that satisfies $l\oplus1$ constraints if such a solution exists.
\end{col_nm}
\junk{
\begin{proof} 
Also this proof is by contradiction and we included in our technical report~\cite{tech_report}.
\end{proof} 
}
\begin{proof}
Based on the Theorem~\ref{th1}, NM provides \textit{the optimal solution} if it exists. Moreover, NM provides a solution which satisfies $l$ constraints if it exists by pruning all links which do not satisfy them during \textit{pre-processing phase} (Algorithm~\ref{buildNbWithPath}, line 9)  Hence, we only need to show in the $l\oplus1$ case that NM always provides a solution that satisfies the path constraint if it exists.
We prove this by contradiction. Assume NM provides a path which does not satisfy the requested path constraint. In this case, the destination node cannot appear in the last neighborhood owing to the path constraint violation (Algorithm~\ref{buildNbWithPath}, line 19). Hence, NM cannot provide this solution. That is a contradiction, which confirms our original thesis.
\end{proof}


\section{Performance evaluation}
\label{implementation}
In this section, we establish the practicality of NM by evaluating  its performance in two complementary scenarios: 
(i) within the management plane, during a virtual network creation when a path is sought, and (ii) within the data plane in traffic steering solution when maintaining a given SLO in a virtual link.
In particular, we compare the performance of NM when applied to the VNE management plane problem, as a link embedding phase. We also analyze the impact of NM during the lifetime of a virtual network and a single NFV chain of middleboxes (virtual nodes) by assessing  the physical network utilization and the energy consumption, when adapting the data path to external physical network state changes.

\begin{figure*}[t!]
\centering

\begin{subfigure}[b]{0.85\textwidth}
\centering
\includegraphics[width=1\linewidth]{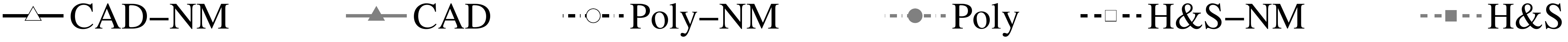}
\end{subfigure}

\begin{subfigure}[b]{0.245\textwidth}
\centering
\includegraphics[width=1\linewidth]{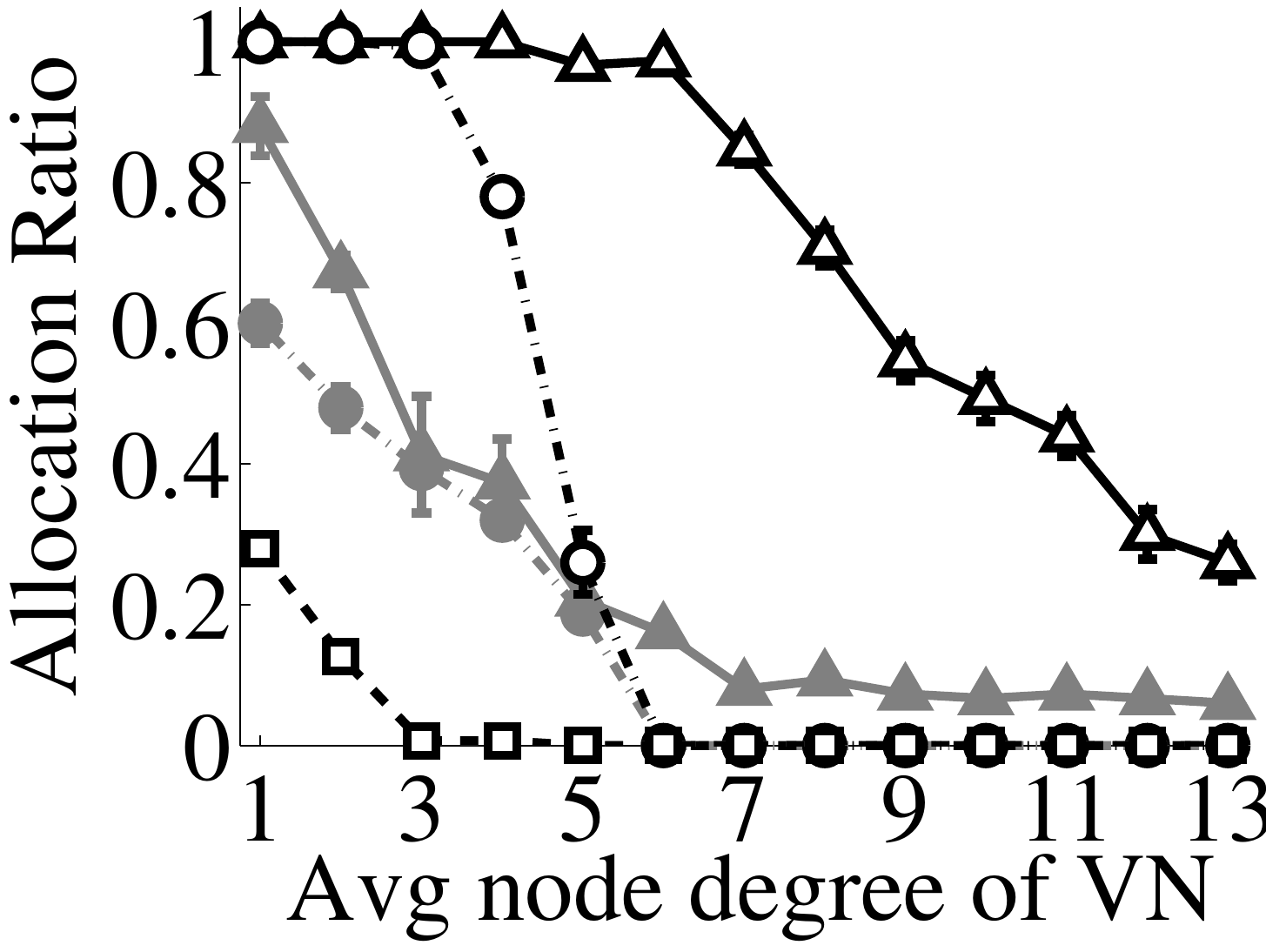}
\caption{}
\label{vne_all}
\end{subfigure}
\begin{subfigure}[b]{0.245\textwidth}
\centering
\includegraphics[width=1\linewidth]{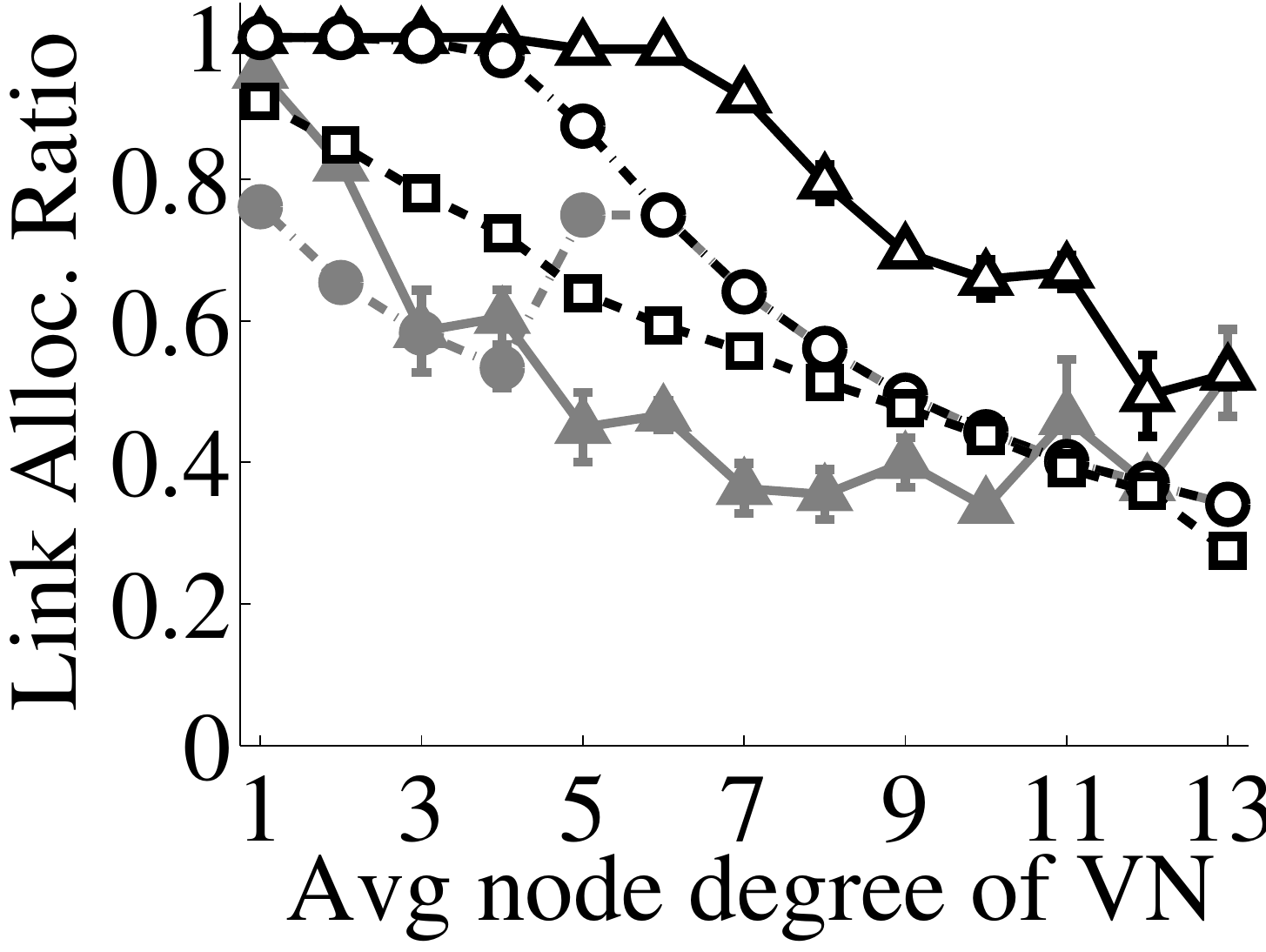}
\caption{}
\label{vne_link}
\end{subfigure}
\begin{subfigure}[b]{0.245\textwidth}
\centering
\includegraphics[width=1\linewidth]{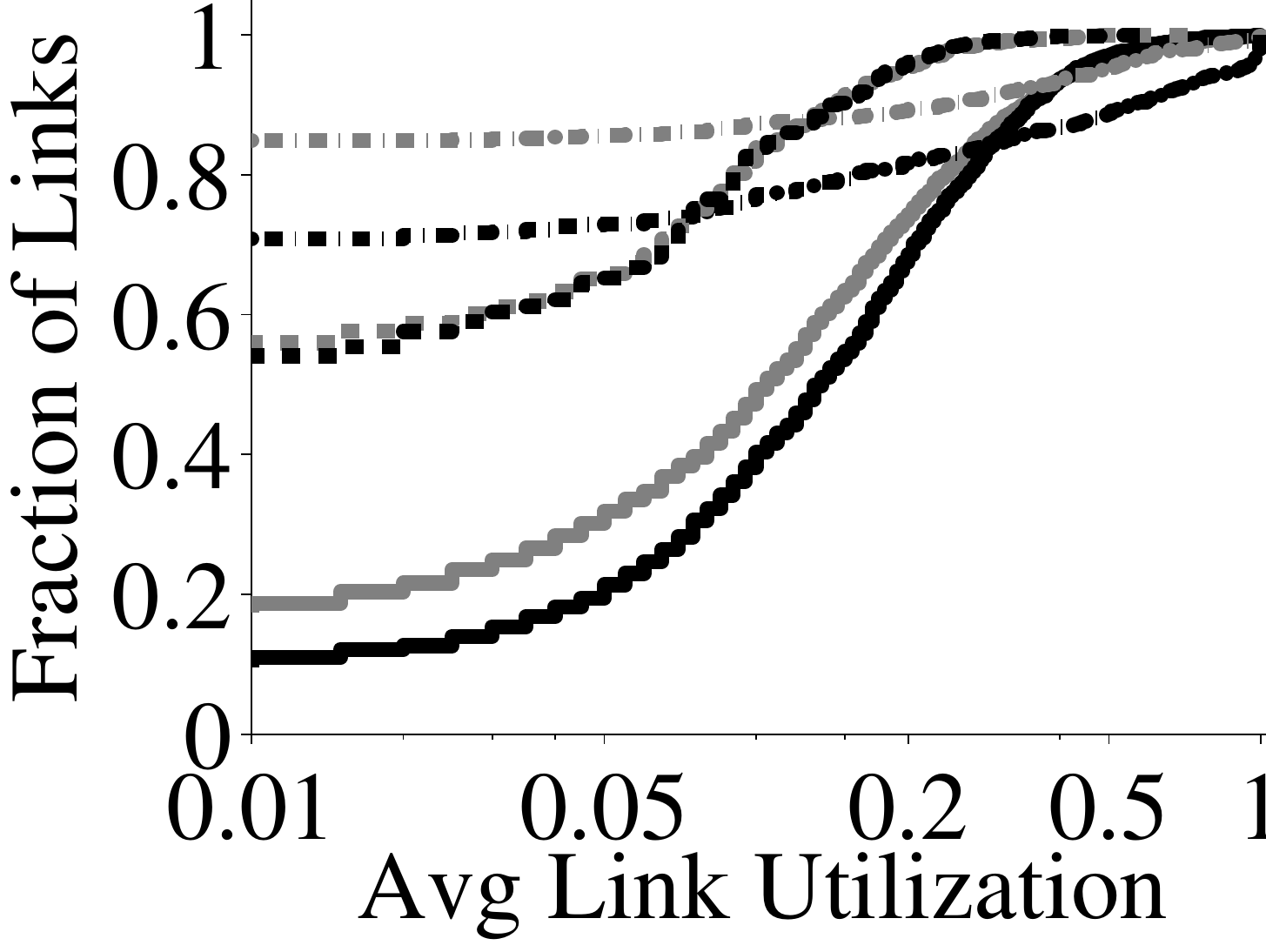}
\caption{}
\label{net_utilization_1}
\end{subfigure}
\begin{subfigure}[b]{0.245\textwidth}
\centering
\includegraphics[width=1\linewidth]{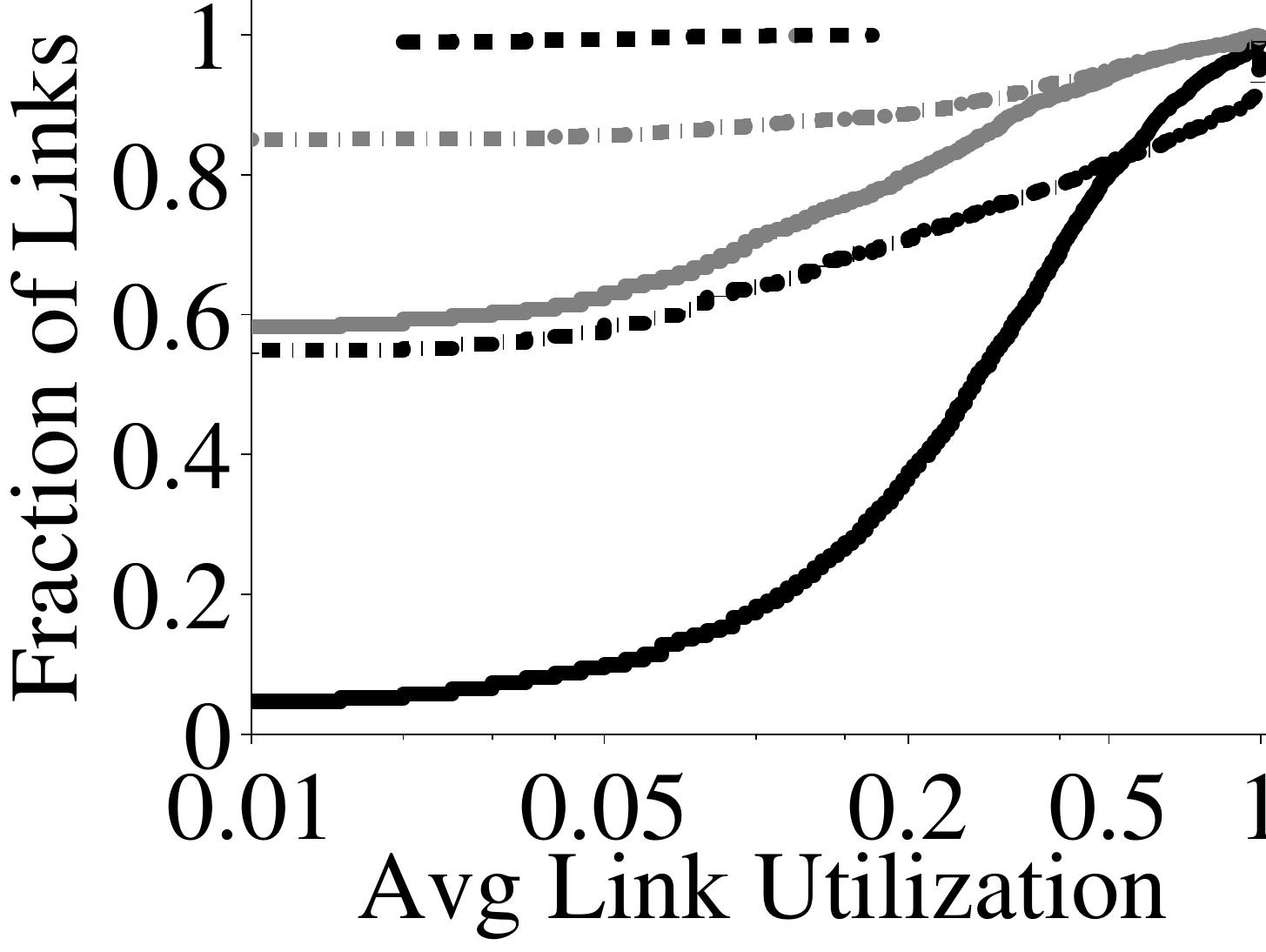}
\caption{}
\label{net_utilization_5}
\end{subfigure}

\caption{\footnotesize{Results obtained with a physical network of $100$ nodes following Waxman connectivity model: on-demand path embedding with NM improves any distributed VNE algorithm utilizing pre-computed $k$-shortest paths.  (a) VN Allocation ratio and (b) link allocation ratio improve when a node embedding phase leaves room for such improvement. Link utilization for (c) linear (NFV chain) and (d) random (with avg node degree of VN = 4) VN topologies is higher for NM-based path embedding.}}
\label{cad_gen}
\vspace{-1mm}
\end{figure*}

\begin{figure*}[t!]
\centering
\begin{subfigure}[b]{1\textwidth}
\centering
\includegraphics[width=1\linewidth]{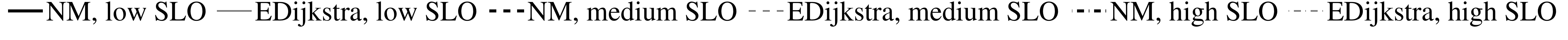}
\end{subfigure}
\hspace{-4mm}
\begin{subfigure}[b]{0.245\textwidth}
\centering
\includegraphics[bb=7 0 381 301,clip=true, width=1\linewidth]{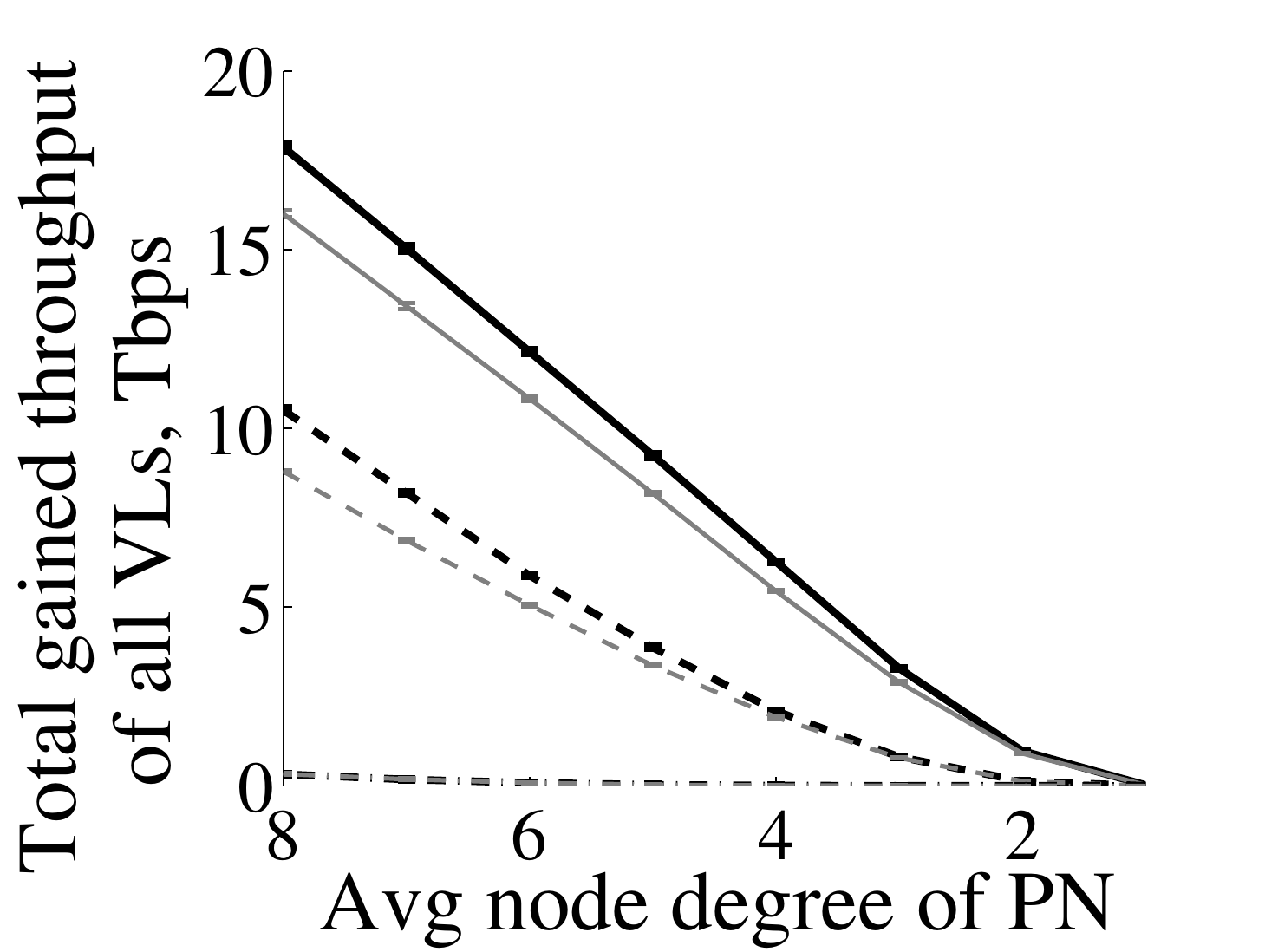}
\caption{}
\label{total}
\end{subfigure}
~
\hspace{-4mm}
\begin{subfigure}[b]{0.245\textwidth}
\centering
\includegraphics[bb=7 0 381 301,clip=true, width=1\linewidth]{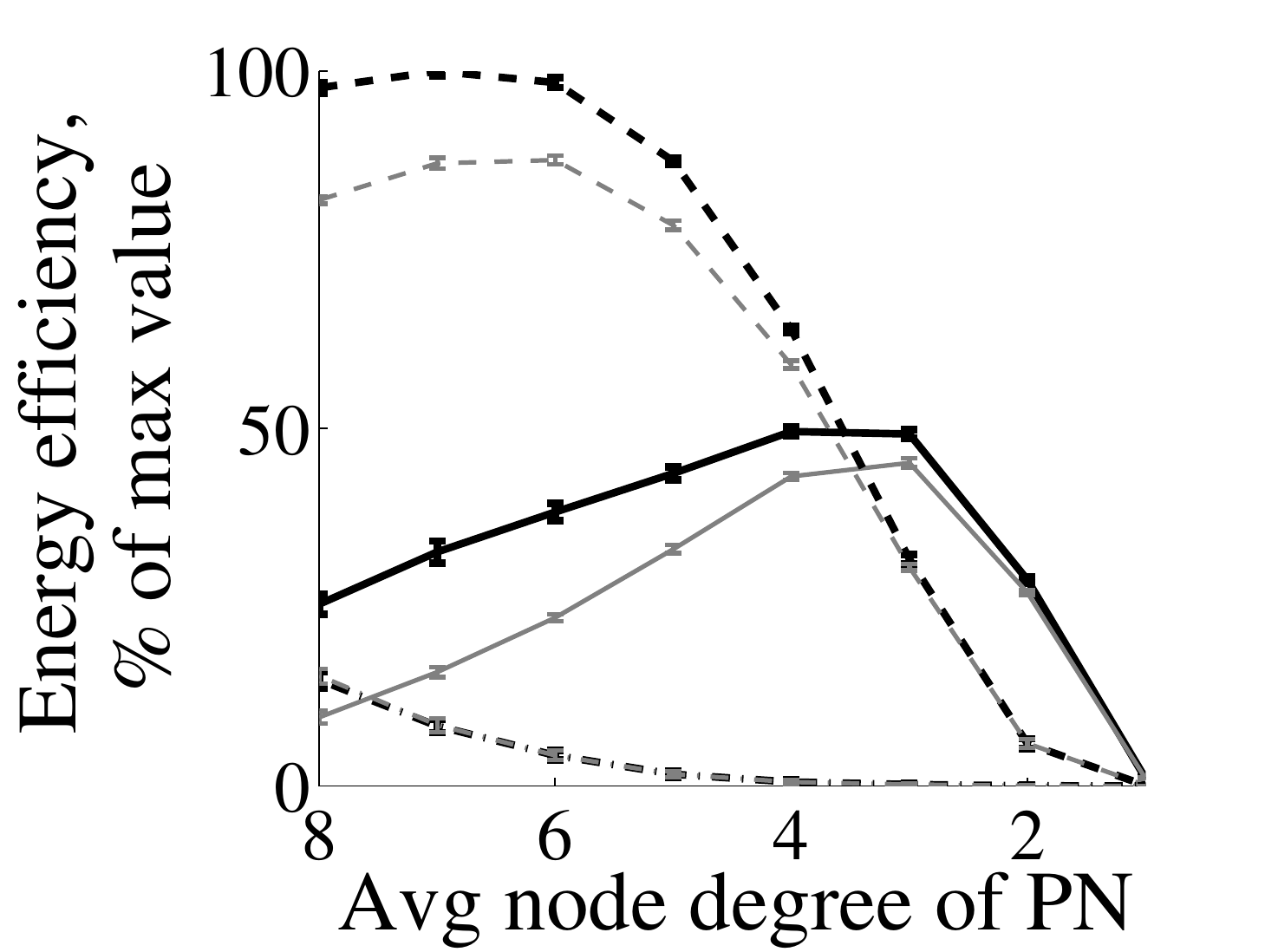}
\caption{}
\label{energy}
\end{subfigure}
~
\hspace{-4mm}
\begin{subfigure}[b]{0.245\textwidth}
\centering
\includegraphics[bb=7 0 381 301,clip=true, width=1\linewidth]{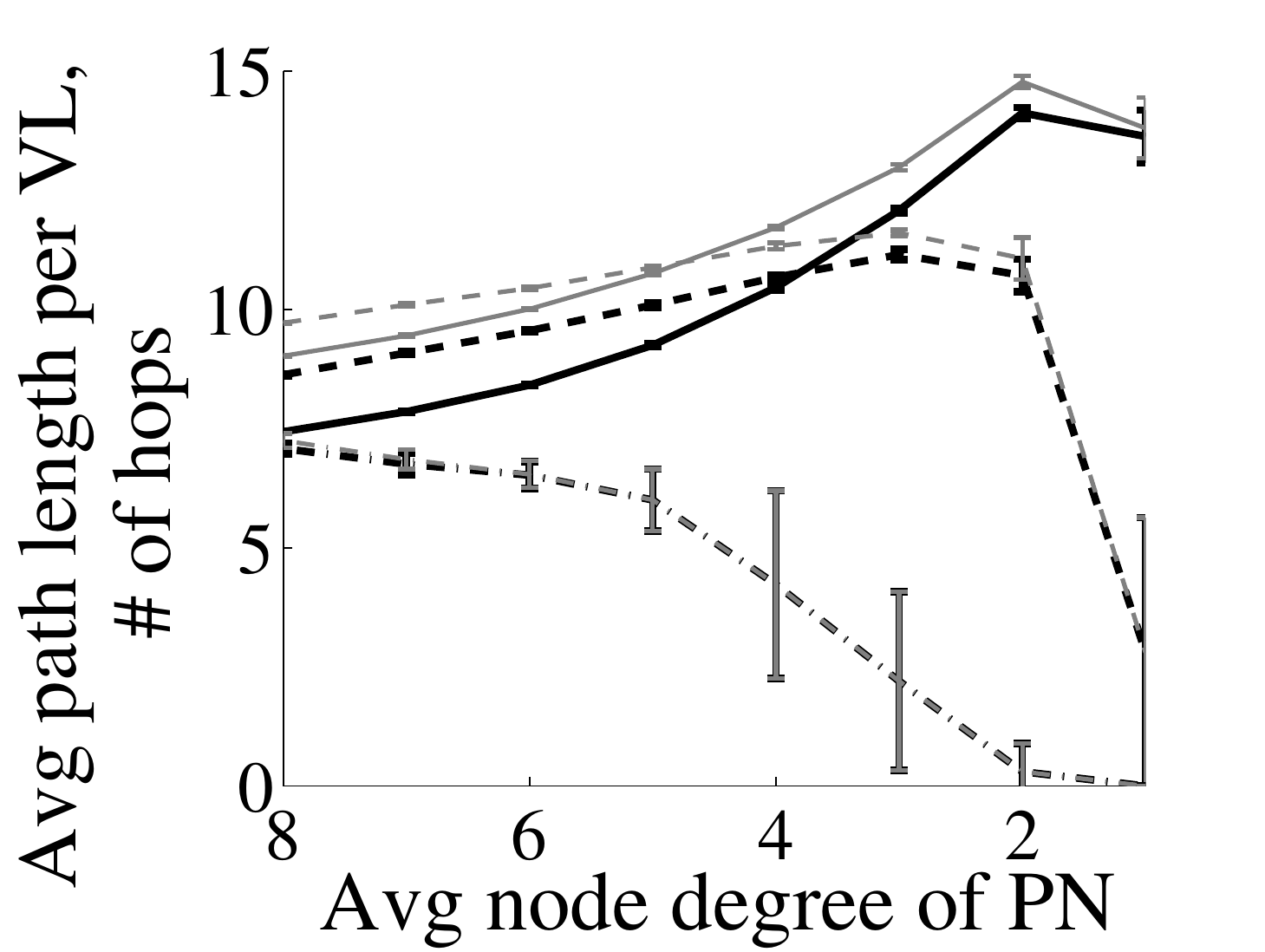}
\caption{}
\label{length}
\end{subfigure}
~
\hspace{-4mm}
\begin{subfigure}[b]{0.245\textwidth}
\centering
\includegraphics[bb=7 0 381 301,clip=true, width=1\linewidth]{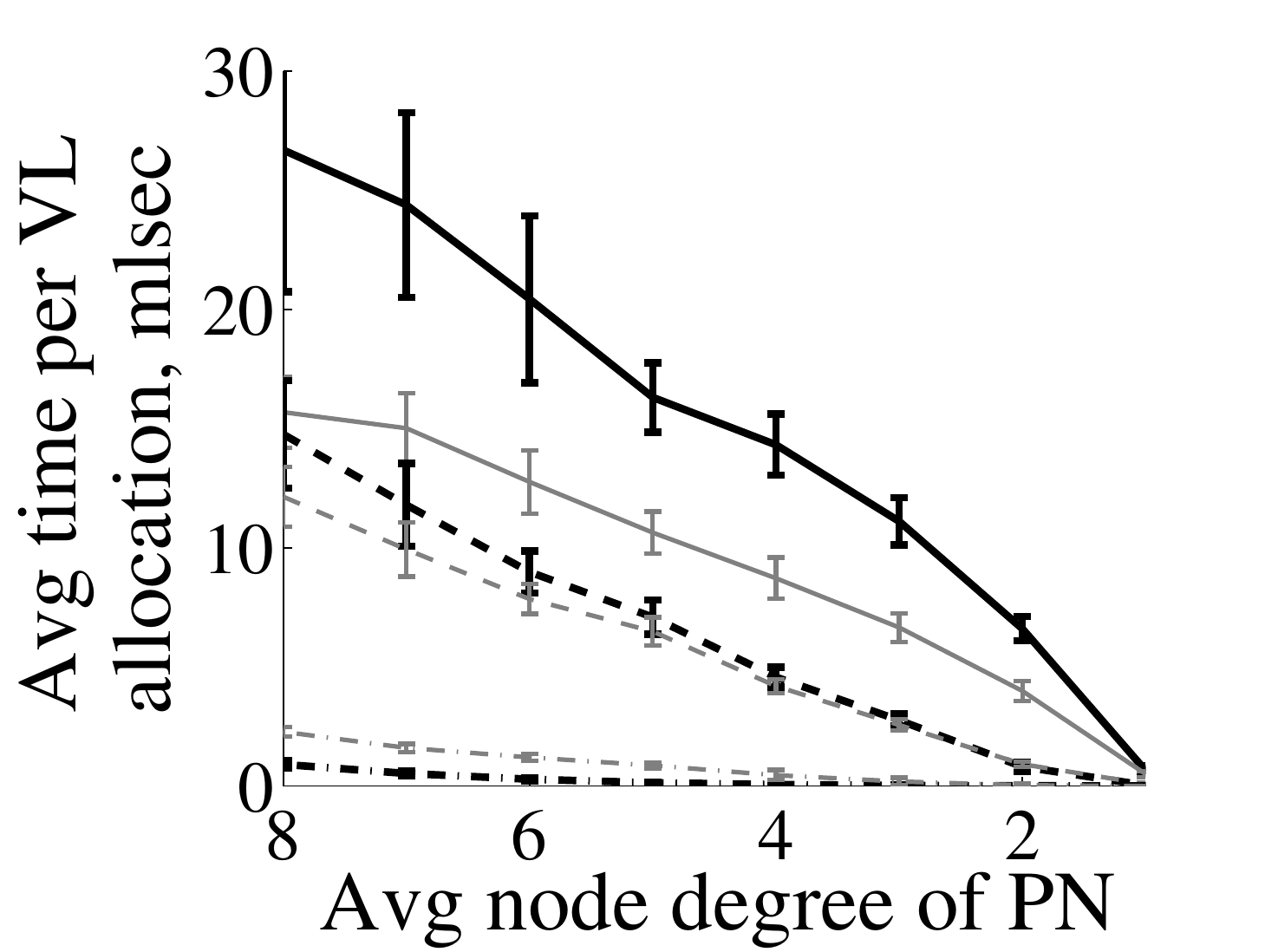}
\caption{}
\label{time}
\end{subfigure}
\caption{\footnotesize{Performance analysis of NM versus EDijkstra on Waxman topologies in terms of: (a) total gained throughput; (b) energy efficiency; (c) average path length; and (d) average time per VL allocation.}}
\label{topo_barabasi}
\vspace{-1mm}
\end{figure*}

\begin{figure*}[t!]
\centering
\begin{subfigure}[b]{1\textwidth}
\centering
\includegraphics[width=1\linewidth]{legend.pdf}
\end{subfigure}
\hspace{-4mm}
\begin{subfigure}[b]{0.245\textwidth}
\includegraphics[bb=7 0 381 301,clip=true, width=1.0\linewidth]{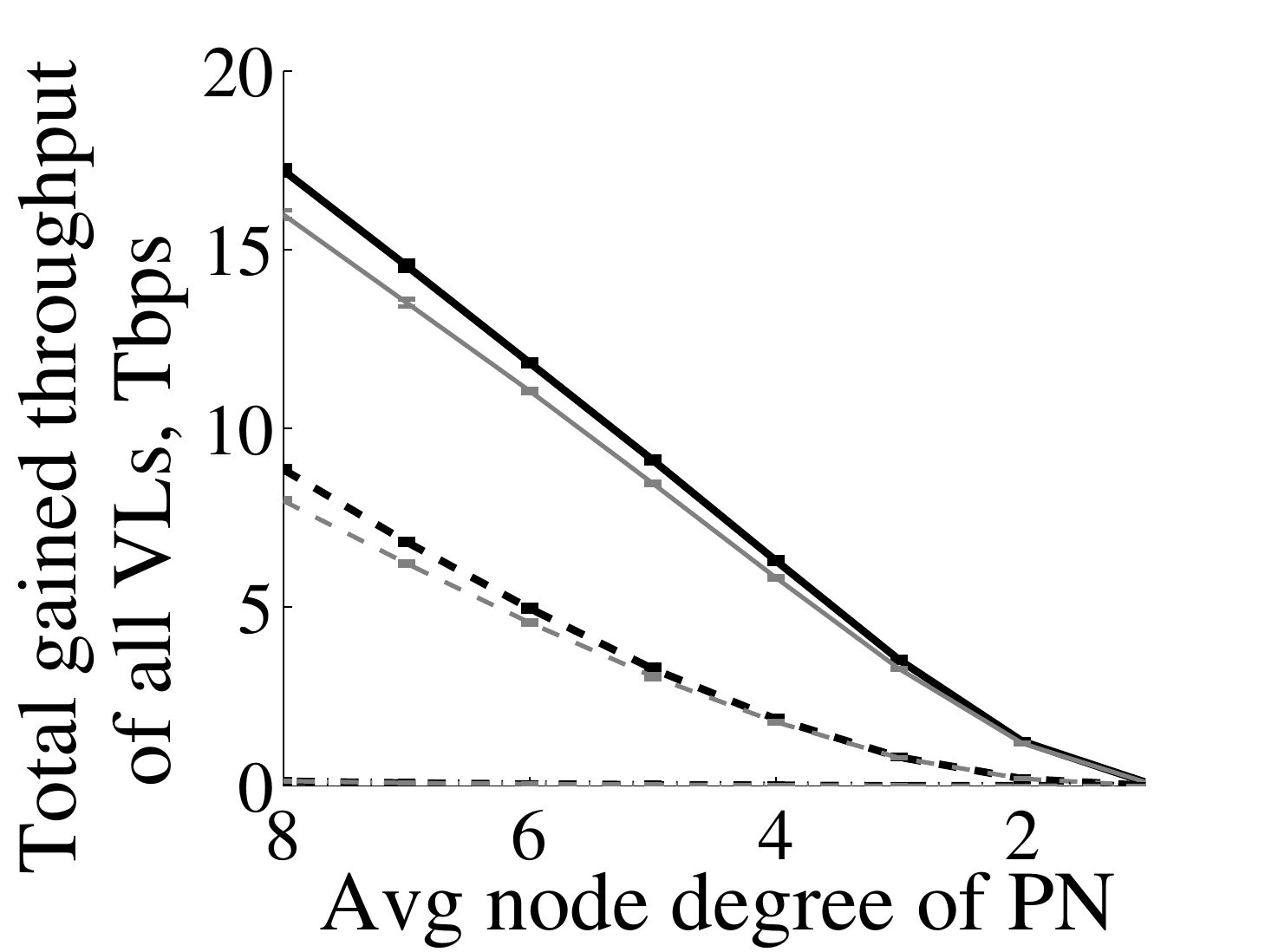}
\caption{}
\label{total_b}
\end{subfigure}
~
\hspace{-4mm}
\begin{subfigure}[b]{0.245\textwidth}
\includegraphics[bb=7 0 381 301,clip=true, width=1.0\linewidth]{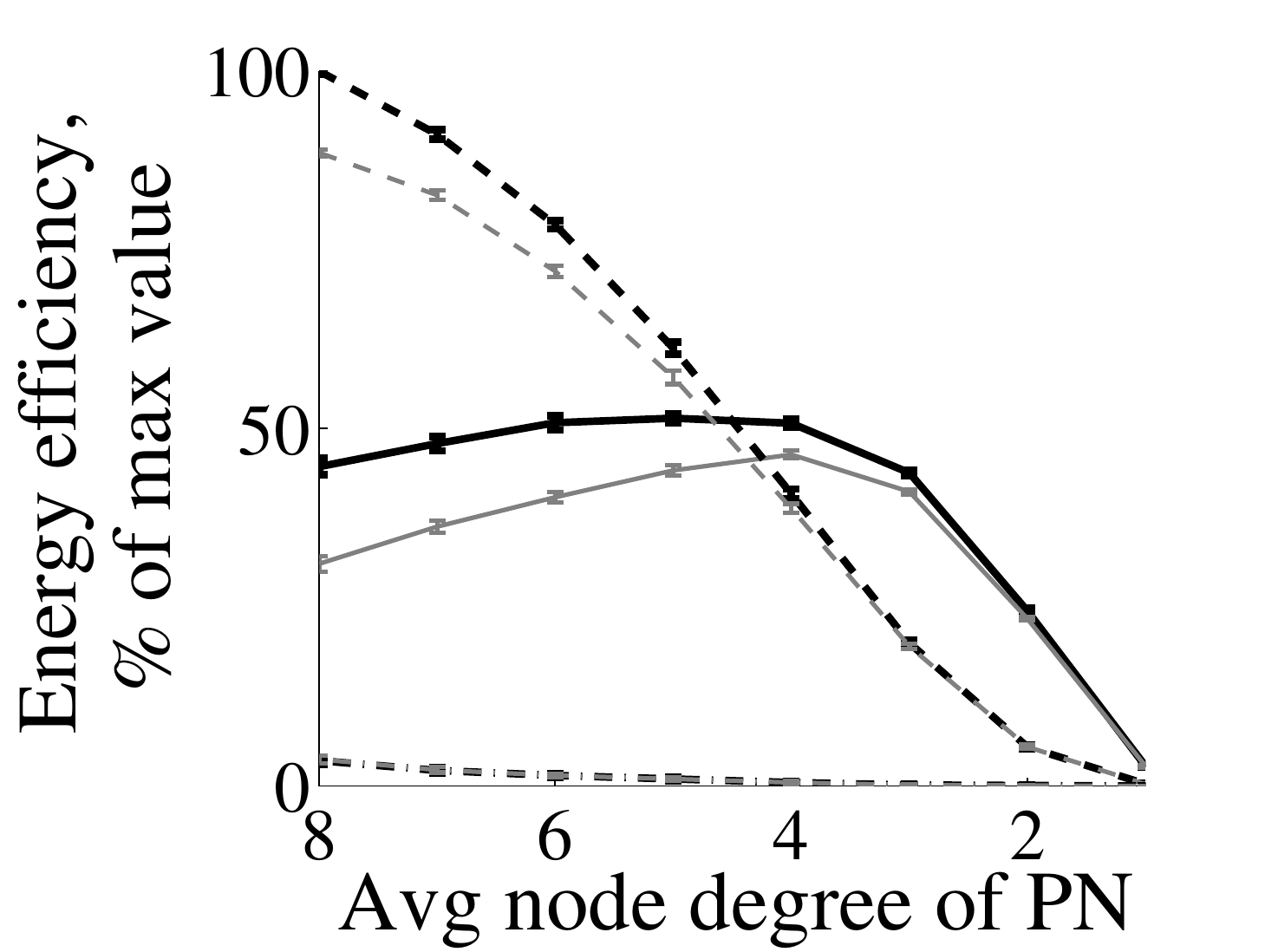}
\caption{}
\label{energy_b}
\end{subfigure}
~
\hspace{-4mm}
\begin{subfigure}[b]{0.245\textwidth}
\includegraphics[bb=7 0 381 301,clip=true, width=1.0\linewidth]{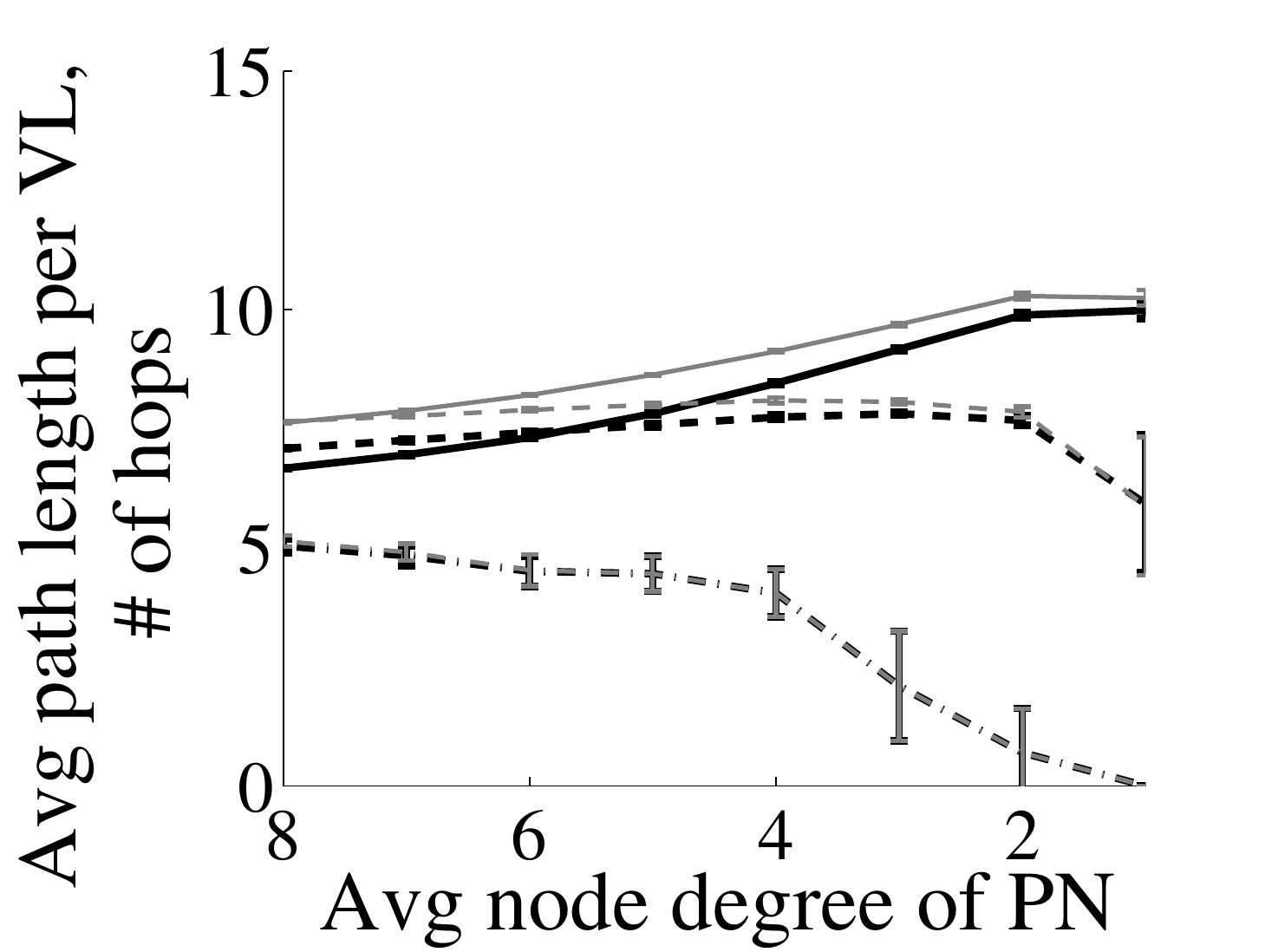}
\caption{}
\label{length_b}
\end{subfigure}
~
\hspace{-4mm}
\begin{subfigure}[b]{0.245\textwidth}
\includegraphics[bb=7 0 381 301,clip=true, width=1.0\linewidth]{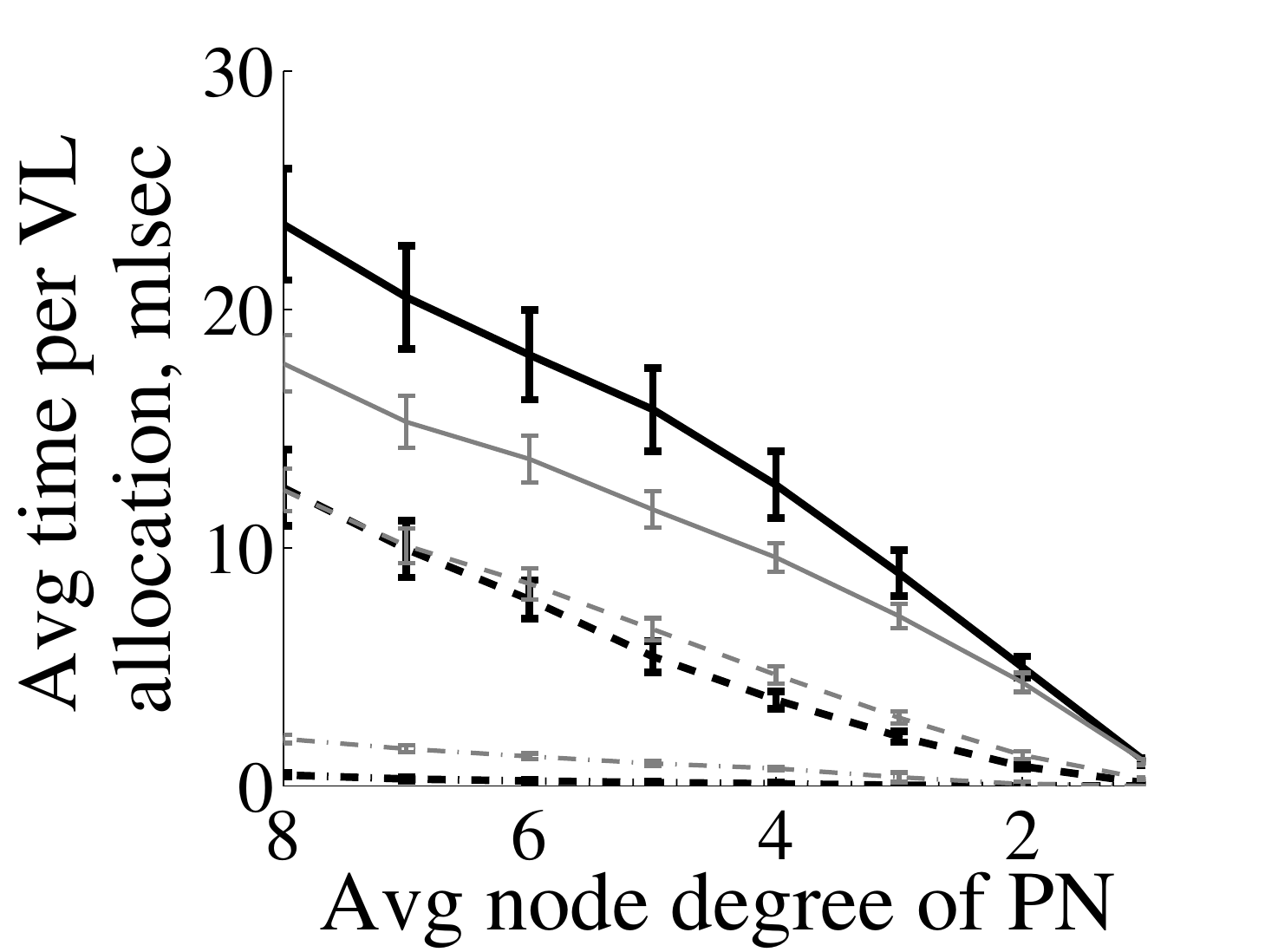}
\caption{}
\label{time_b}
\end{subfigure}
\caption{\footnotesize{Performance analysis of NM versus EDijkstra  on Barabasi-Albert topologies in terms of: (a) total gained throughput; (b) energy efficiency; (c) average path length; and (d) average time per VL allocation.}}
\label{topo_barabasi}
\vspace{-4mm}
\end{figure*}

\noindent
{\bf Simulation Settings.}
For our simulations, we used a machine with an \textit{Intel(R) Xeon(R)} processor with CPU $2.1$ GHz and $1$GB RAM, and running the Ubuntu OS GNU/Linux $x86\_64$. 
We use the BRITE~\cite{brite} topology generator to create our physical and virtual networks. Our results are consistent across physical networks which follow both Waxman and Barabasi-Albert models~\cite{waxman}~\cite{internet}. All our results show $95\%$ confidence interval, and our randomness lays in both the (virtual) link constraints and in the physical network topology. 

To assess the impact of NM on the virtual network embedding, we include here the results obtained with a physical network of $100$ nodes following Waxman connectivity model, where each physical node has $200$ CPU units and $200$ bandwidth units, and each virtual node and link have up to $20$ units of CPU and bandwidth, respectively. We attempt to embed a pool of $15$ VN requests with $14$ virtual nodes and random virtual topologies, so that we could vary the number of virtual links from $1$ (the VN has linear topology) to $13$ (VN is a fully connected topology).~\footnote{Using a dataset of $8$ years of $64,000$ real VN embedding requests to the Emulab~\cite{emulab} virtual network testbed, it has been shown in~\cite{nodeembedding} that VNs have size of $14$ nodes on average.}  
%

To evaluate the impact of NM on data plane traffic steering applications, we simulate instead requests for remapping constrained virtual links. This scenario suits also traffic steering applications in a dynamic NFV chain reallocation when a new policy is declared. In particular, on a physical network topology of $10,000$ nodes, where each physical link has bandwidth uniformly distributed  between $1$ and $9$ Gbps, we attempt to find an optimal path for $1000$ random pairs $<src, dst>$, where for each pair we allocate as many virtual links as possible with the same demands.
We denote as low, medium and high bandwidth constraints, 1, 4 and 7 Gbps, respectively, which represent approximately $10\%$, $45\%$ and $80\%$ of the maximum physical link capacity, while with high, medium and low delay constraints we indicate $400\%$, $250\%$, and of $80\%$  of the maximum physical link delay, respectively.
\textcolor{black}{An additional simulation involves using a fixed average physical node degree equal to 4 (which is common for the Internet~\cite{internet}), different bandwidth ($BW$) constraints considered in the case of SLO demands while varying the delay constraint. We request a delay constraint from $400\%$ to $50\%$ of the maximum link delay.}

\noindent
{\bf Management plane evaluation metrics.}  To demonstrate the advantages of using NM as a path embedding solution, we compared three representative VNE distributed algorithms and replaced  their proposed link embedding solutions  with our NM. 
We have chosen Hub-and-Spoke (H\&S)~\cite{HS} as to our knowledge it is the first distributed VNE solution. In ~\cite{HS} physical nodes elect the host of the hub virtual node, and then the host for the spokes virtual nodes. Then a virtual link (VL) embedding phase finds the shortest path among the hosts. If at least a VL constraint is violated or a virtual node cannot be hosted, the entire VN request is rejected. 
We also compare against PolyViNE (Poly)~\cite{polyvine}, the first policy-based distributed VNE solution. Physical nodes belonging to different infrastructure providers partition the VN and attempt to embed the largest possible VN partitions, and then create virtual links among the winner physical hosts.
Finally, we compare against a Consensus-based Auction mechanism (CAD)~\cite{nodeembedding}, the first policy-based distributed VNE approximation algorithm with convergence and optimality guarantees. The link embedding of these algorithms runs a $k$-shortest path with $k=3$ for Waxman and $k=1$ for Barabasi-Albert topologies shown to be within optimal range of $k$.

In this simulation scenario we have tested the potential revenue loss by using a suboptimal distributed VNE algorithm by specifying the fraction of VN request accepted over the VN requesed (allocation ratio), how many virtual links (VL) were accepted over the VL requested (link allocation ratio)  and to what extent physical links were utilized (link utilization). Note that in all other link embedding solutions, all shortest paths are pre-computed, while NM computes them dynamically.

\noindent
{\bf Data plane evaluation metrics.}
To evaluate the performance of NM for data plane traffic steering solutions we compare NM with the Extended Dijkstra algorithm or $k$-shortest path algorithm with $k=1$, where both algorithms find a path on-demand. 
We compare NM across four metrics: total gained throughput of all reallocated VLs, energy efficiency, average path length per VL, and average (convergence) time to a stable path. We approximate the energy efficiency with the ratio between unused nodes and the total gained throughput:
\vspace{-2mm}
\begin{equation}
\mbox{Energy Efficiency}=\frac{N-N_{used}}{N} \cdot bw_{total}
\label{energy_eq}
\end{equation} 
where $N$ is the total number of physical nodes, $N_{used}$ is the number of physical used nodes to maintain all VLs, and $bw_{total}$ is the number of VLs multiplied by their $bw$. 

\noindent
{\bf Results Summary.} 
During the virtual network formation, we found that using NM to dynamically search a loop-free physical path to embed a constrained virtual link may significantly increase  the virtual network allocation ratio (Figure~\ref{vne_all}), resulting in higher physical link utilization (Figures~\ref{net_utilization_1} and~\ref{net_utilization_5}). 
During the virtual network lifetime instead, we show how embedding a path with NM is beneficial in terms of optimal path length, network utilization and energy efficiency. Such advantages come at different intensity levels, depending on the severity of the requested constraints, and bring along with them some interesting tradeoffs. 

\noindent
{\bf Path embedding during virtual networks creation.}  Except when the node embedding phase leaves no room for allocation ratio improvements (see $e.g.$, the Hub\&Spoke heuristic), NM brings higher link embedding acceptance rates (see Figure~\ref{vne_all}). This is because, when a lower-cost alternative path is available, NM finds it (see Figure~\ref{vne_link}). We conclude that seeking on-demand low-cost paths brings higher VN allocation ratios as well as higher overall network utilization, hence 
\textcolor{black}{making the Cloud and infrastructure provider services more attractive to customers} 
(see Figures~\ref{net_utilization_1} and~\ref{net_utilization_5}). 
\noindent
The price we pay to dynamically find the optimal loop-free hosting physical path is higher response time: on-demand path computation cannot be as fast as pre-computing a suboptimal path. 

\begin{figure*}[t!]
\centering
\begin{subfigure}[b]{1\textwidth}
\centering
\includegraphics[width=0.98\linewidth]{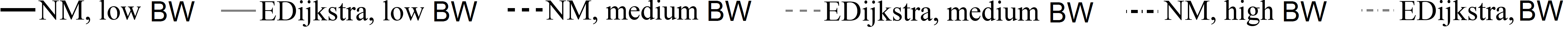}
\end{subfigure}
\hspace{-4mm}
\begin{subfigure}[b]{0.245\textwidth}
\includegraphics[bb=7 0 381 301,clip=true, width=1.0\linewidth]{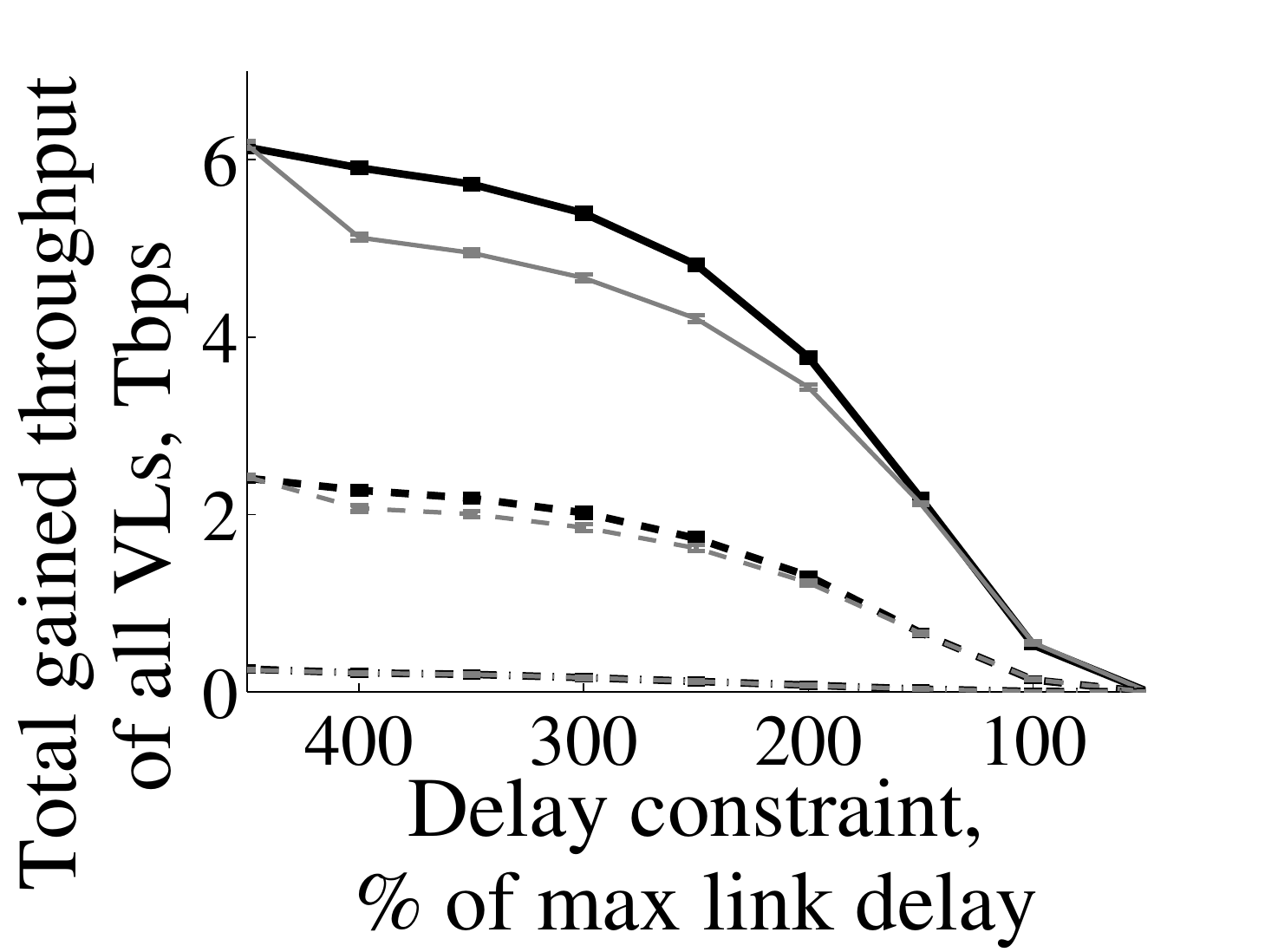}
\caption{}
\label{a_total}
\end{subfigure}
~
\hspace{-4mm}
\begin{subfigure}[b]{0.245\textwidth}
\includegraphics[bb=7 0 381 301,clip=true, width=1.0\linewidth]{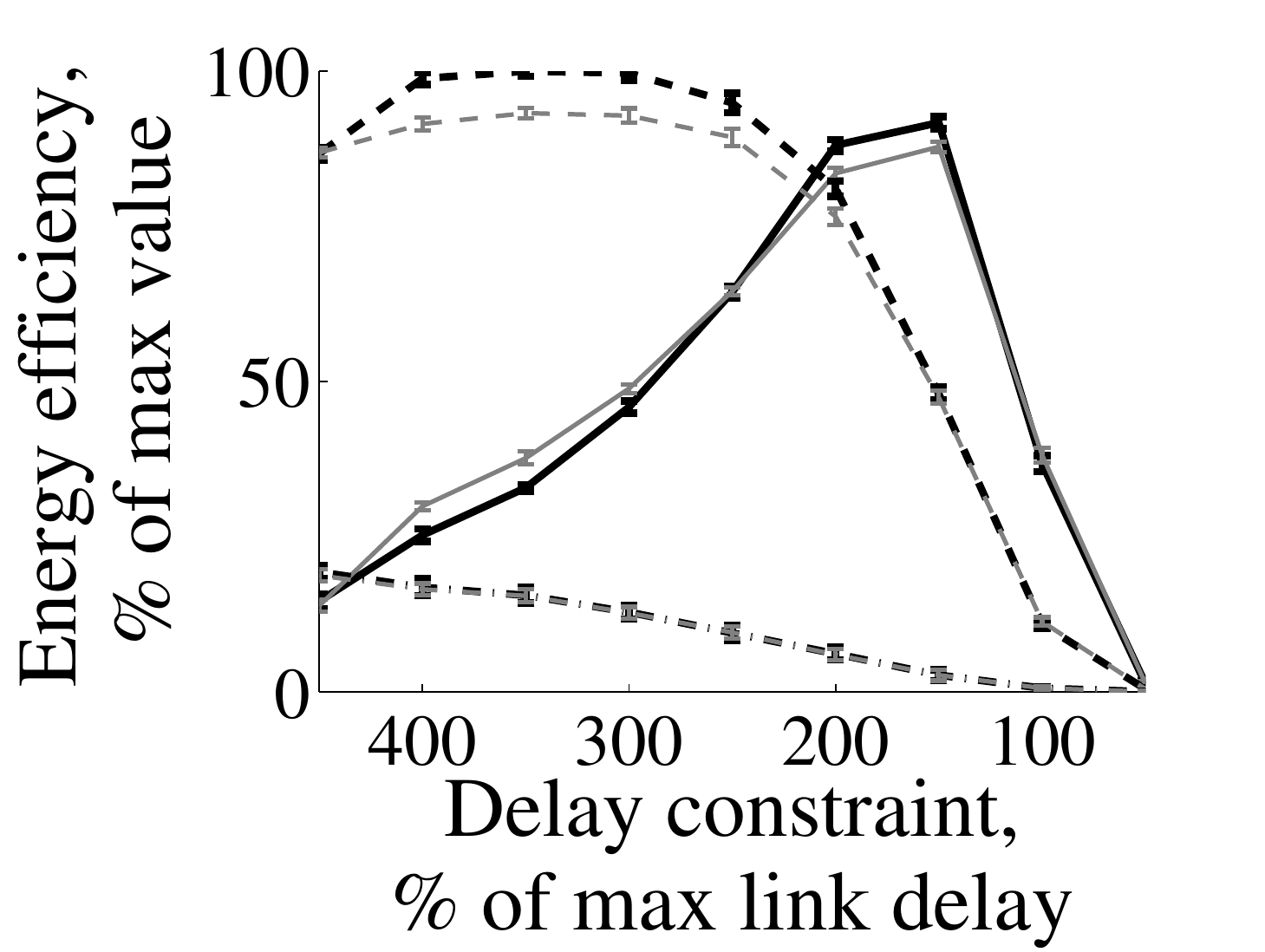}
\caption{}
\label{a_energy}
\end{subfigure}
~
\hspace{-4mm}
\begin{subfigure}[b]{0.245\textwidth}
\includegraphics[bb=7 0 381 301,clip=true, width=1.0\linewidth]{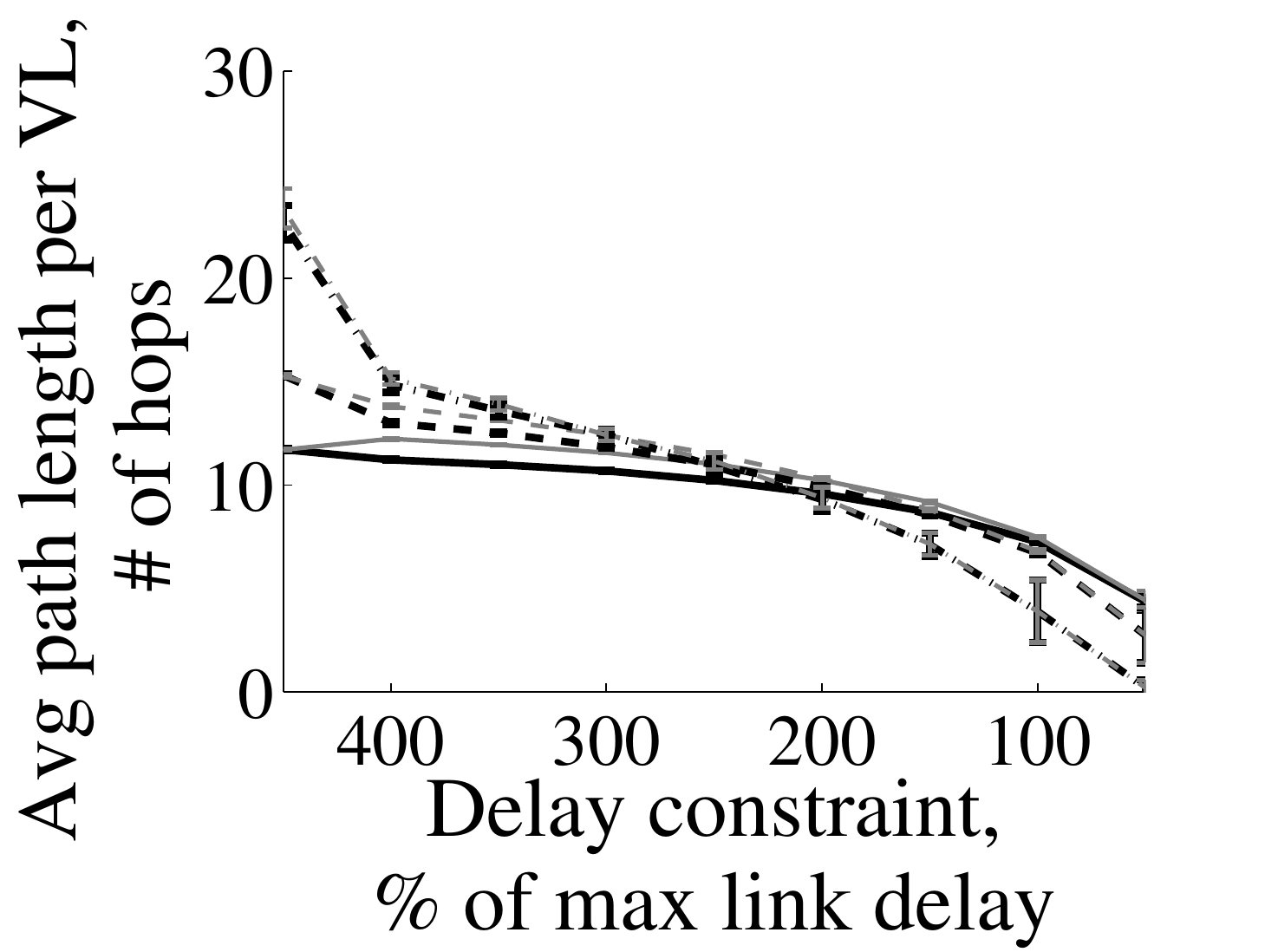}
\caption{}
\label{a_length}
\end{subfigure}
~
\hspace{-4mm}
\begin{subfigure}[b]{0.245\textwidth}
\includegraphics[bb=7 0 381 301,clip=true, width=1.0\linewidth]{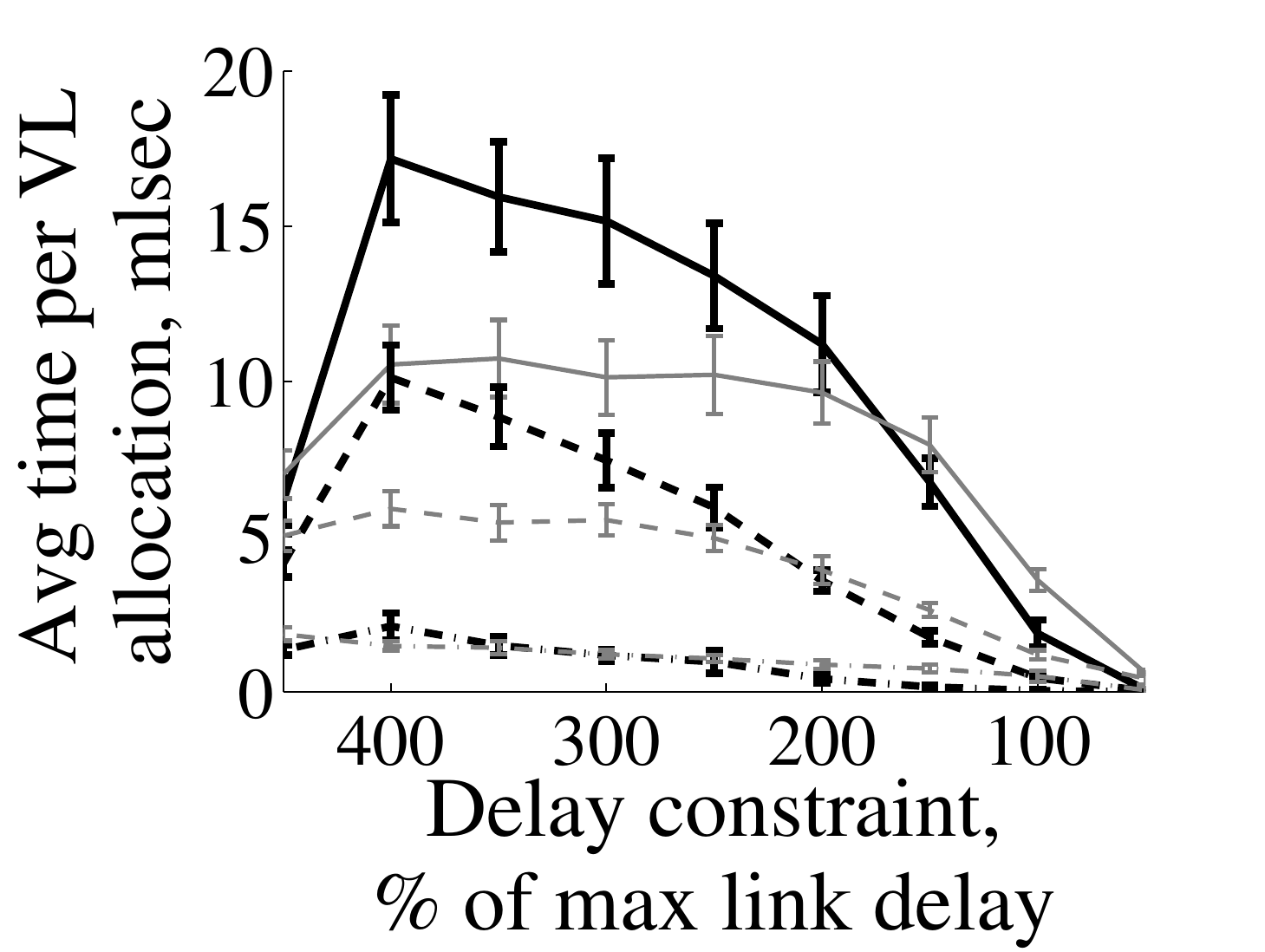}
\caption{}
\label{a_time}
\end{subfigure}
\caption{\footnotesize{SLO constraints correlation analysis of NM versus EDijkstra on Waxman topologies in terms of: (a) total gained throughput; (b) energy efficiency; (c) average path length; and (d) average time per VL allocation.}}
\label{topo_barabasi}
\vspace{-1mm}
\end{figure*}

\begin{figure*}[t!]
\centering
\begin{subfigure}[b]{1\textwidth}
\centering
\includegraphics[width=0.98\linewidth]{a_legend.png}
\end{subfigure}
\hspace{-4mm}
\begin{subfigure}[b]{0.245\textwidth}
\includegraphics[bb=7 0 381 301,clip=true, width=1.0\linewidth]{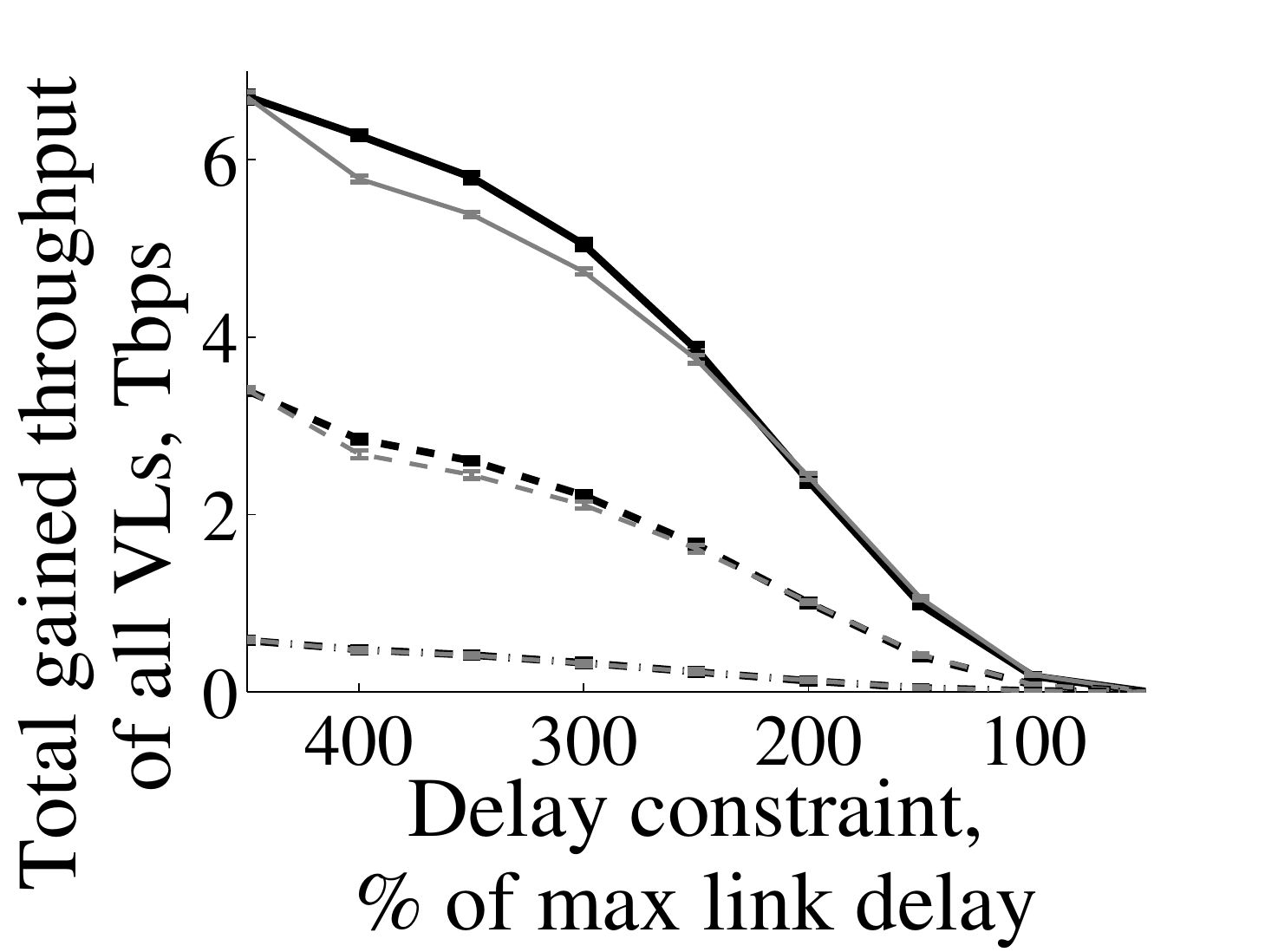}
\caption{}
\label{a_total_b}
\end{subfigure}
~
\hspace{-4mm}
\begin{subfigure}[b]{0.245\textwidth}
\includegraphics[bb=7 0 381 301,clip=true, width=1.0\linewidth]{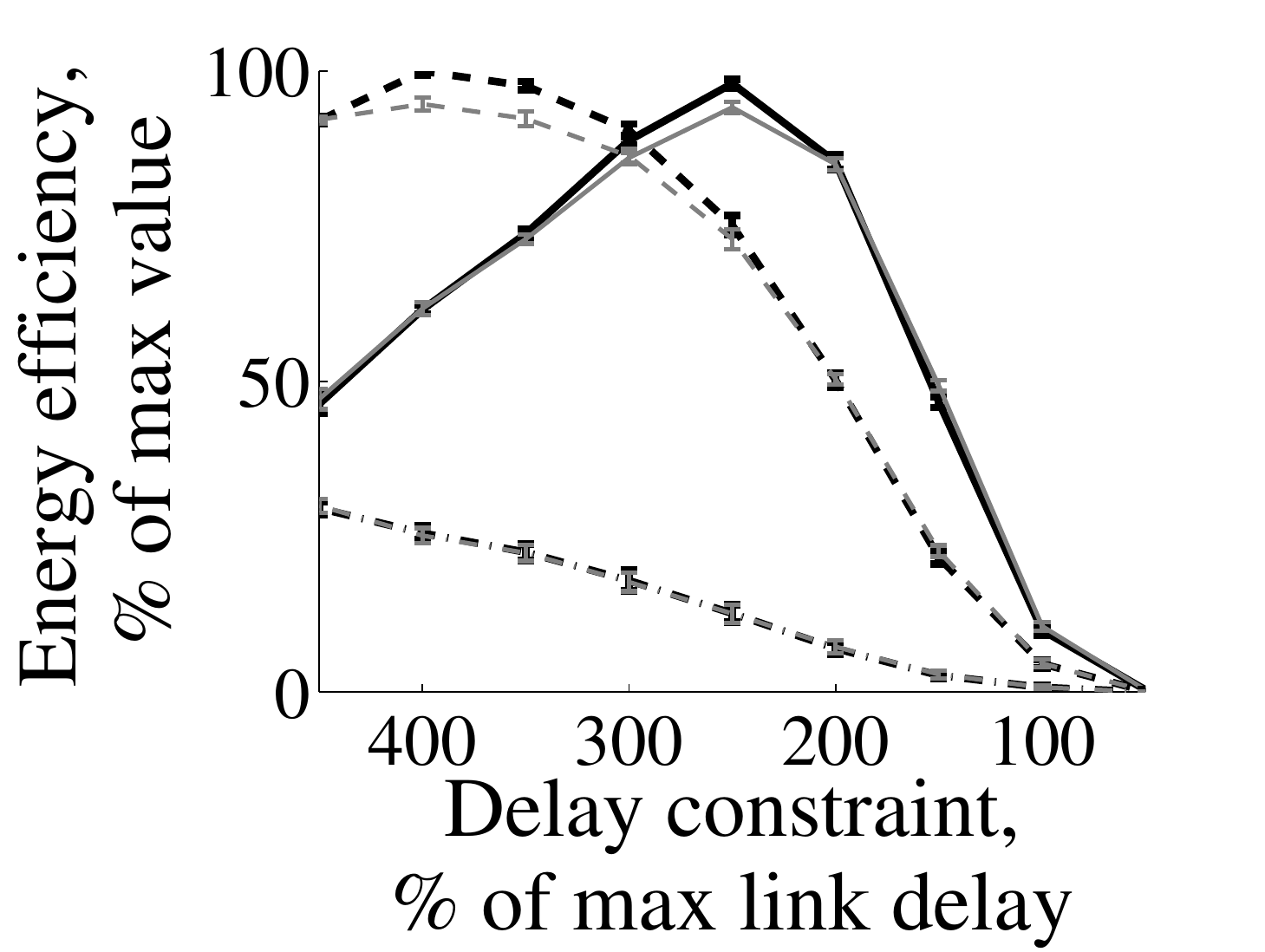}
\caption{}
\label{a_energy_b}
\end{subfigure}
~
\hspace{-4mm}
\begin{subfigure}[b]{0.245\textwidth}
\includegraphics[bb=7 0 381 301,clip=true, width=1.0\linewidth]{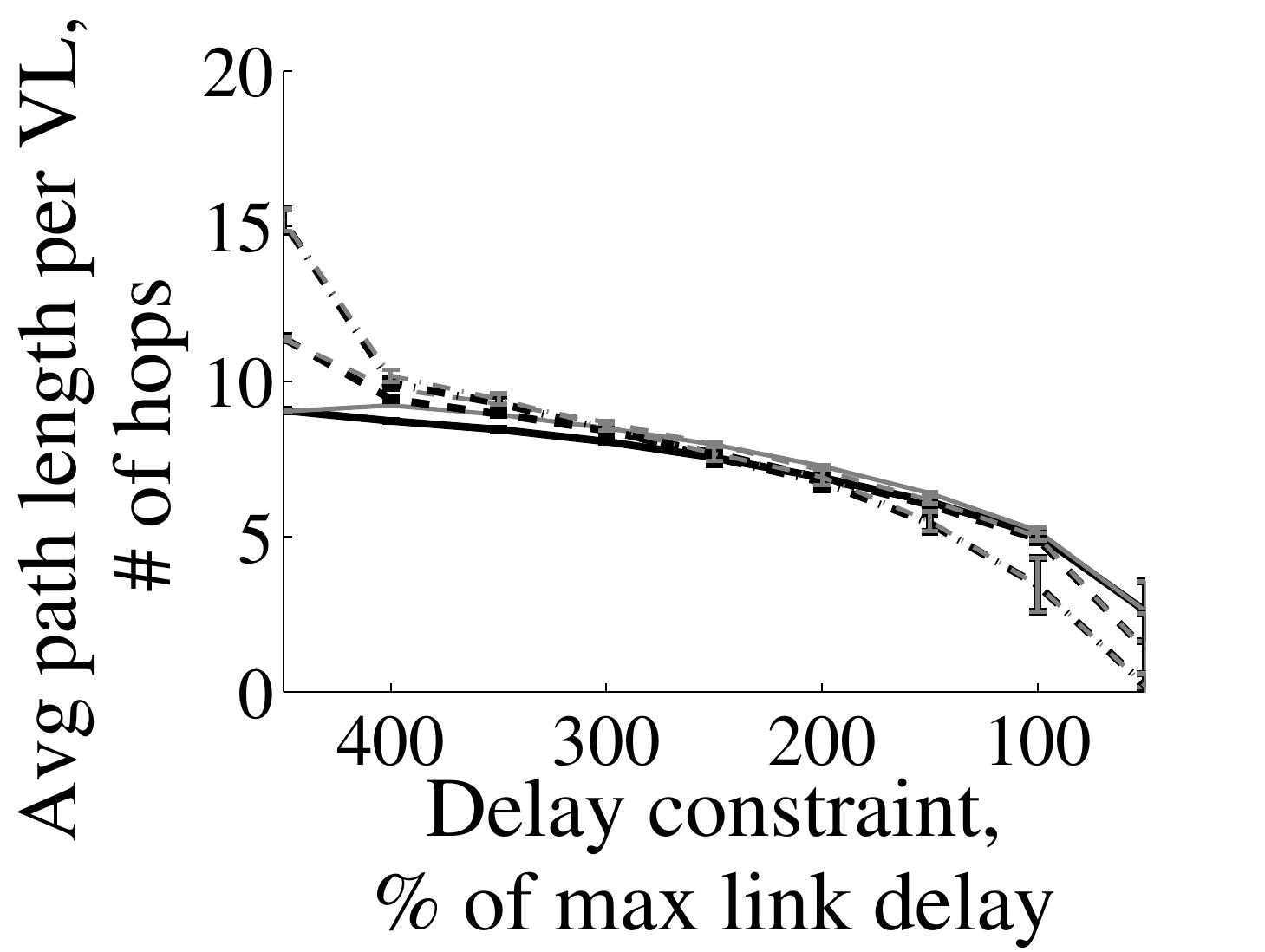}
\caption{}
\label{a_length_b}
\end{subfigure}
~
\hspace{-4mm}
\begin{subfigure}[b]{0.245\textwidth}
\includegraphics[bb=7 0 381 301,clip=true, width=1.0\linewidth]{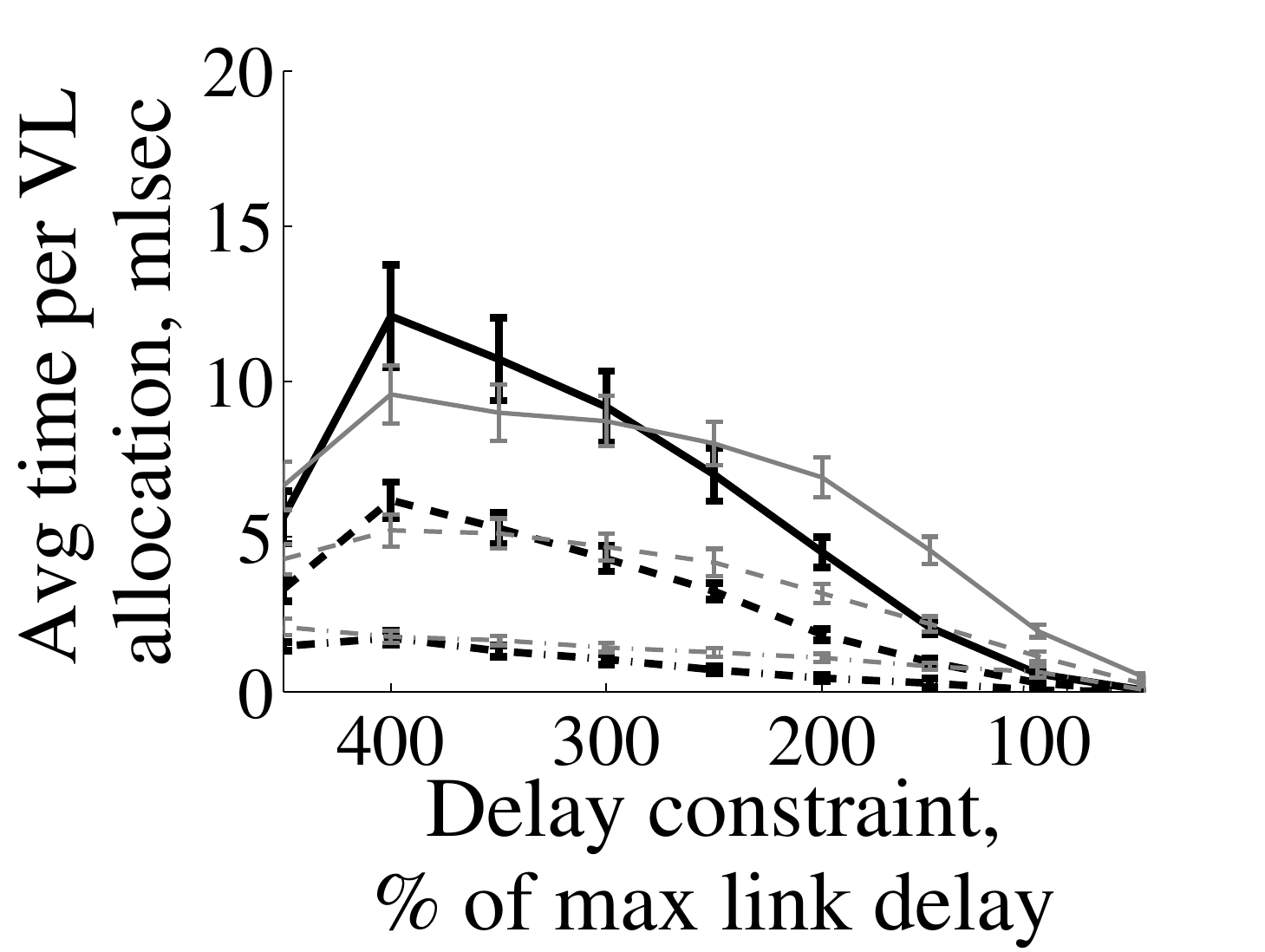}
\caption{}
\label{a_time_b}
\end{subfigure}
\caption{\footnotesize{SLO constraints correlation analysis of NM versus EDijkstra on Barabasi-Albert topologies in terms of: (a) total gained throughput; (b) energy efficiency; (c) average path length; and (d) average time per VL allocation.}}
\label{topo_barabasi}
\vspace{-4mm}
\end{figure*}

\noindent
{\bf NM shows benefits in physical network utilization and energy efficiency.}
The next set of results analyze the performance of our NM during path maintenance in comparison with EDijkstra under physical network topologies with a different average node degree, as well as with different SLO constraints.

In Figure$~\ref{total}$ and$~\ref{total_b}$, we show how the total gained throughput of all reallocated VLs changes when multiple physical links become available (the average physical node degree increases). 
This dependence is an affine function with a negative coefficient: the maximum possible total gained throughput also increases linearly with the available physical links. Moreover, we found how the total gained throughput is higher for low SLO constraints. This is not surprising as the path constraints requirement becomes harder to satisfy.

Note also how the over provisioning of the network utilization obtained by our NM (black lines) outperforms EDijkstra (gray lines). The main difference between these two algorithms in the $l\oplus 1$ case is a path length optimization. 
This result demonstrates how reducing path over provisioning in traffic steering techniques may significantly increase the physical network utilization. 
As expected, this gain degrades as constraints become harder to satisfy, and decreases with the node connectivity, as less physical path choices are available to map our virtual links. 

\noindent
{\bf NM shows energy efficiency gains.}
In Figures$~\ref{energy}$ and $~\ref{energy_b}$, we illustrate the impact of physical path over provisioning on the energy efficiency of the physical network. Even in the medium SLO constraints case, NM gains are up to $\approx$20\%; while with low SLO constraints NM gains are up to 150\% in  energy efficiency for Waxman topology and slightly lower for Barabasi-Albert topology, w.r.t. a traffic steering technique run using the EDijkstra. Even in this case, the gain degrades when the SLO constraints increase, and with the shortage of available physical links. 

More surprising is the observed trade-off between energy gain and SLO constraints: we found that the peak of energy efficiency is obtained when the paths have neither lenient nor high SLO constraints. This is because, when the SLO constraints are low, both algorithms allocate many longer paths which, in turn utilize a lot of physical resources. On the other hand, in the case of high SLO constraints, the available physical paths are limited, and the total throughput gain remains small.

\noindent
{\bf  Average path length tradeoff and over provisioning.}
We further investigate the reasons why we observed such gains in both throughput and energy efficiency w.r.t. EDijksra (Figures$~\ref{length}$,$~\ref{time}$,$~\ref{length_b}$, and$~\ref{time_b}$).
In particular, observing Figures$~\ref{length}$ and $~\ref{length_b}$ we note that there are almost 2 hops difference in the average path length between NM and EDijkstra for the high average node degree for Waxman topology and slightly lower for Barabasi-Albert topology, whereas for the medium SLO constraints this difference is reduced to circa one hop for Waxman topology and it is even lower for Barabasi-Albert topology. Finally, when the SLO constraints are high, there is no significant physical path length difference.
This is confirmed by Figures$~\ref{time}$ and$~\ref{time_b}$, where we show that NM has an higher average time per embedded path than EDijkstra for the low and medium SLO constraints, which is in line with NM's quadratic time complexity of  $l\oplus 1$ case.

To understand why the average path length changes with the constraint severity, note how the longer is an end-to-end physical path, the lower is the probability that
the entire path satisfies the SLO constraints. On the other hand, the higher the number of hops, the higher is the number of candidates paths, and so the higher is the probability of finding one which satisfies these constraints. This explains the trade-offs in average path length behavior observed in Figures$~\ref{length}$ and$~\ref{length_b}$.

\noindent
{\bf Correlation between SLO constraints.}
To better understand the correlation between link (bandwidth) and path (delay) constraints, additional results are as follows. Figures$~\ref{a_total}$ and$~\ref{a_total_b}$ show a logarithmic function - with delay increasing the total gained throughput growths logarithmically. Again, NM gains up to 1 Tbps throughput with respect to what EDijkstra produces when we set the bandwidth to low and delay to high constraints (6 Tbps versus 5 Tbps). This translates into an extra $20\%$ of network utilization for Waxman topology and lesser difference for Barabasi-Albert topology. 

Similar trade-offs can be observed in energy efficiency w.r.t. previous scenario (Figures$~\ref{a_energy}$ and $~\ref{a_energy_b}$) where NM shows $10\%$ and $5\%$ higher maximum energy efficiency than EDijkstra for Waxman and Barabasi-Albert topologies, respectively. Despite the performance in the medium bandwidth case, NM always shows better or equal energy efficiency. Note that in all simulations where the bandwidth is set to low values and delay constraint is set to high and medium values, we observe a trade off between total throughput (higher for NM) and energy efficiency (higher for EDijkstra). However, NM is never worse for both total gained throughput and energy efficiency. Further, Figures$~\ref{a_length}$ and $~\ref{a_length_b}$ again confirm optimality of NM solution. However, this time we may see logarithmic dependency from delay constraint is explained by unlikeliness of having longer path for low delay (high SLO constraint). Figures$~\ref{a_time}$ and$~\ref{a_time_b}$ also agree with NM's quadratic time complexity of  $l\oplus 1$ case.

\section{Conclusion}
\label{conclusion}

In this paper we defined the Virtual Path Embedding problem, $i.e.$, the problem of embedding a virtual path on a physical or logical constrained loop-free path minimizing the network over provisioning.
To solve this problem, we proposed a novel on-demand path embedding scheme $viz.$, ``Neighborhoods Method" (NM). NM is suitable for both management plane mechanisms such as virtual network embedding and network functions virtualization, as well as for data plane mechanisms such as traffic steering satisfying SLOs.  In particular, we show how NM provides on-demand SLO virtual path guarantees while reducing expensive physical network over provisioning, by considering multiple link and a single path constraints in polynomial time. 

Via Monte-Carlo simulations for a set of diverse topology scenarios, we also show that our NM approach can lead to gains of up to $20\%$ in network utilization during the virtual network creation phase within both the management and the data planes, and up to $150\%$ in energy efficiency during the virtual network lifetime within the data plane, with respect to existing embedding solutions based on the multi-constrained $k$-shortest path or Extended Dijkstra.

\end{document}